%% file: main.tex
\documentclass[conference]{IEEEtran}
\IEEEoverridecommandlockouts
\pdfoutput=1
\usepackage{cite}
\usepackage{amsmath,amssymb,amsfonts}
\usepackage{graphicx}
\usepackage{textcomp}
\usepackage{subcaption}
\usepackage[table,xcdraw]{xcolor}
\usepackage[]{algorithm2e}
\def\BibTeX{{\rm B\kern-.05em{\sc i\kern-.025em b}\kern-.08em
    T\kern-.1667em\lower.7ex\hbox{E}\kern-.125emX}}
\usepackage{multirow}
\usepackage{colortbl}
\usepackage{hhline}
\usepackage{tikz}
\usepackage{amssymb}
\usepackage{pifont}
\usepackage{diagbox}
\usepackage{url}
\usepackage{verbatim}
\usepackage{soul}
\def\checkmark{\tikz\fill[scale=0.3](0,.35) -- (.25,0) -- (1,.7) -- (.25,.15) -- cycle;}

\usepackage{booktabs} 
\usepackage[symbol]{footmisc}

\usepackage{slashbox}

\usepackage{makecell}

\input{commands}

\begin{document}

\title{Performance Characterization of using Quantization for DNN Inference on Edge Devices: Extended Version}

\author{
	Hyunho Ahn\IEEEauthorrefmark{1}, Tian Chen\IEEEauthorrefmark{1}, Nawras Alnaasan, Aamir Shafi, Mustafa Abduljabbar, Hari Subramoni, and \\ Dhabaleswar K. (DK) Panda \\
	The Ohio State University\\
    
    \begin{normalsize}
        \begin{sffamily}
            \{ahn.377, chen.9891, alnaasan.1, shafi.16, abduljabbar.1, subramoni.1, panda.2\}@osu.edu
        \end{sffamily}
    \end{normalsize}

}

\maketitle
\pagestyle{plain}
\begin{abstract}
\input{text/abstract}
\end{abstract}

\begin{IEEEkeywords}
Quantization, Edge, Inference, MLPerf
\end{IEEEkeywords}

\footnote[0]{\IEEEauthorrefmark{1}These authors contributed equally to this work\\}

\input{text/intro}
\input{text/background}

\input{text/quantization.tex}

\input{text/setup}

\input{text/evaluation}

\input{text/related_work}

\input{text/conclusion}
\input{text/acknowledgments.tex}

\bibliographystyle{ieeetr}
\bibliography{bibs/main}

\end{document}

%% file: commands.tex
\usepackage{pifont}

\usepackage{url}

%% file: text/abstract.tex
Quantization is a popular technique used in Deep Neural Networks (DNN) inference to reduce the size of models and improve the overall numerical performance by exploiting native hardware. This paper attempts to conduct an elaborate performance characterization of the benefits of using quantization techniques---mainly FP16/INT8 variants with static and dynamic schemes---using the MLPerf Edge Inference benchmarking methodology. The study is conducted on Intel x86 processors and Raspberry Pi device with ARM processor. The paper uses a number of DNN inference frameworks, including OpenVINO (for Intel CPUs only), TensorFlow Lite (TFLite), ONNX, and PyTorch with MobileNetV2, VGG-19, and DenseNet-121. The single-stream, multi-stream, and offline scenarios of the MLPerf Edge Inference benchmarks are used for measuring latency and throughput in our experiments. Our evaluation reveals that OpenVINO and TFLite are the most optimized frameworks for Intel CPUs and Raspberry Pi device, respectively. We observe no loss in accuracy except for the static quantization techniques. We also observed the benefits of using quantization for these optimized frameworks. For example, INT8-based quantized models deliver $3.3\times$ and $4\times$ better performance over FP32 using OpenVINO on Intel CPU and TFLite on Raspberry Pi device, respectively, for the MLPerf offline scenario. To the best of our knowledge, this paper is the first one that presents a unique characterization study characterizing the impact of quantization for a range of DNN inference frameworks---including OpenVINO, TFLite, PyTorch, and ONNX---on Intel x86 processors and Raspberry Pi device with ARM processor using the MLPerf Edge Inference benchmark methodology.
 

%% file: text/intro.tex
\vspace{-4.0ex}

\section{Introduction}
\label{sec:intro}

The last decade has seen the emergence of Deep Neural Network (DNN) training as an important workload on parallel systems, including High-Performance Computing and Cloud hardware. DNNs have been found to be very useful in many applications, including Computer Vision and Natural Language Processing, due to their high accuracy that is mainly due to the large number of training parameters. While significant successes~\cite{densenet,VGG,mobilenetv2} have been realized in training such large networks, there is relatively less focus on deploying them for inference on edge devices. The deployment of these large models for inference on commodity servers, as well as resource-constrained environments, is vital for successful {\em democratization} of Artificial Intelligence (AI) models. 

A common challenge in deploying large models for inference is the sheer size of these models due to the large number of parameters. One technique to address this is quantization which allows using the lower-precision number format for storing weights and activations during DNN training and inference~\cite{Survey_of_quantization}. This means using formats like INT8, FP16, etc., instead of the default FP32. While quantization has been very successful in DNN training, this paper focuses on inference only.
\input{text/motivation.tex}

\input{text/contribution.tex}

%% file: text/motivation.tex
\subsection{Motivation}
\label{sec:intro:motivation}

The main motivation of this paper is to conduct performance characterization using quantization for DNN inference on edge systems, including Intel x86 systems and Raspberry Pi 4B device equipped with ARM processor. We are interested in quantifying the reduction in sizes of quantized models while also measuring the accuracy of these models---the goal is to reduce the size while not affecting the accuracy. We are also motivated to explore and use the commonly used quantization techniques, including FP16 and INT8 variations. This study is done using a variety of DNN inference frameworks, including OpenVINO~\cite{openvino} (for Intel CPUs only), TensorFlow Lite (TFLite)~\cite{TFLite}, ONNX~\cite{ONNX_Runtime}, and PyTorch~\cite{pytorch} using specialized backends and libraries for the corresponding x86 and ARM processors. The overall goal of using quantization is to: 1) reduce the memory/energy footprint of AI models without losing accuracy and 2) improve numerical performance by exploiting native hardware support for faster arithmetic. We use the benchmarking methodology adopted by the MLPerf Edge Inference benchmarks~\cite{MLPerf}.

%% file: text/contribution.tex
This paper makes the following key contributions:

\begin{itemize}

  \item Explore the use of various quantization techniques---based on INT8/FP16 and static/dynamic strategies---on a range of DNN inference frameworks, including OpenVINO, PyTorch, TFLite, and ONNX.

  \item The performance evaluation is done on Intel CPUs (Cascade Lake and Skylake) and Raspberry Pi 4B equipped with ARM processor.
  
  \item The performance characterization reveals that the size of original models is reduced by a quarter for INT8-based models without losing accuracy. The only exception is when static quantization is utilized, where we witnessed a slight accuracy reduction.
  
  \item The characterization study uses a range of popular AI models---including MobileNetV2~\cite{mobilenetv2}, VGG-19~\cite{VGG}, and DenseNet-121~\cite{densenet}. We found that OpenVINO and TFLite are the most optimized frameworks for Intel CPUs and Raspberry Pi 4B device, respectively. For the MLPerf offline scenario, INT8-based quantized models deliver $3.3\times$ and $4\times$ better performance over FP32 using OpenVINO on Intel CPU and TFLite on Raspberry Pi device, respectively.

  \item The evaluation is done using the MLPerf Edge Inference benchmark and uses the single-stream, multi-stream, and offline scenarios. We also studied the impact of using optimized numerical instructions like Vector Neural Network Instruction (VNNI)~\cite{VNNI} provided by the Cascade Lake processors.
  
\end{itemize}

{\em To the best of our knowledge, this paper presents a unique characterization study that studies the impact of quantization for a range of DNN inference frameworks---including OpenVINO, TFLite, PyTorch, and ONNX---on Intel x86 processors and Raspberry Pi device with ARM processor using the MLPerf Edge Inference benchmark methodology.}

Rest of the paper is organized as follows. Section~\ref{sec:bgnd} presents background on DNN inference frameworks the MLPerf Edge Inference benchmark. Section~\ref{sec:quant} reviews important concepts related to quantization and provides an overview of our approach to quantizing models for OpenVINO, PyTorch, TFLite, and ONNX. The experimental setup for our characterization study is provided in Section~\ref{sec:setup} that is followed by the detailed evaluation and analysis in Section~\ref{sec:evaluation}. Section~\ref{sec:related_work} presents related work and the paper is concluded in Section~\ref{sec:conclusion}.

%% file: text/background.tex
\section{Background}
\label{sec:bgnd}

\subsection{Deep Learning Frameworks on Edge Devices}

Deep Learning (DL) frameworks provide a high-level interface and building blocks for designing, training, and validating Deep Neural Networks (DNNs) on a wide range of devices. There is a plethora of ML/DL frameworks such as TensorFlow~\cite{tensorflow}, PyTorch~\cite{pytorch}, CoreML~\cite{coreML}, ONNX~\cite{ONNX_Runtime}, OpenVINO~\cite{openvino}.
Each of these frameworks differs in terms of purpose, performance, model API, and hardware compatibility. Some frameworks are designed for a specific hardware architecture, like CoreML, which is exclusively used for Apple devices. Other frameworks like OpenVINO and TensorFlow Lite (TFLite)~\cite{TFLite} are more focused on providing an efficient and portable solution for model inference on devices that have limited memory and computing resources. 

One solution to address the limitations of edge devices is the quantization of DL models to reduce the size and compute requirements for performing inference tasks. Furthermore, several low-level libraries can be used to accelerate the performance of edge devices. 
For instance, the ArmNN library~\cite{armnn} bridges the gap between the DL framework and underlying architectures by increasing the efficiency of the Arm Cortex-A CPUs and Arm Mali GPUs. 
ONNX supports similar libraries like NVIDIA TensorRT~\cite{tensorRT} and Intel oneDNN~\cite{oneDNN}. 
Intel also provides its optimized TensorFlow version for Intel CPUs~\cite{intel_tensorflow}, which uses oneDNN to fully utilize the Advanced Vector eXtensions (AVX) instruction set.

For our experiments, we select four representative frameworks that support model quantization: 1) PyTorch, which allows the training, quantization, and deployment of models within the same framework, 2) TFLite, which is the optimized TensorFlow runtime for edge devices, 3) ONNX, which offers great flexibility in translating models from/to other DL frameworks, and 4) OpenVINO, an Intel developed framework which is integrated with several Intel acceleration libraries.

\subsection{MLPerf Inference Benchmark}

The MLPerf Inference Benchmark Suite~\cite{MLPerf} is a standard machine learning (ML) benchmark suite that prescribes a set of rules and best practices to fairly evaluate the inference performance of ML hardware. 
It spans multiple ML models and tasks in the Computer Vision and Natural Language Processing domains, including image classification, object detection, medical imaging, speech-to-text, translation, etc. Each task and model are well-defined to ensure the reproducibility and accessibility of the benchmarks. 
An MLPerf Inference submission system consists of System Under Test (SUT), Load Generator (LoadGen), Accuracy Script, and Data Set unit. SUT includes the hardware, architecture, and software used in the inference. SUT should follow Model-equivalence rules, which provide a complete list of disallowed and allowed techniques in benchmarking. These rules are in place to help submitters efficiently reimplement models on various architectures. 
The LoadGen is a traffic generator that loads the SUT and measures performance. It produces the query traffic according to the rules of each scenario. 
MLPerf identifies four inference scenarios that represent many critical inference applications in real-life use cases: the single-stream, multi-stream, server and offline scenarios. Among the server senario is not required in edge benchmark. We conduct our experience in the remaining three scenarios. In each scenario, the LoadGen process generates inference requests in a particular pattern. In the single-stream and multi-stream scenarios, the LoadGen sends the next query as soon as SUT completes the previous query. In offline scenario, LoadGen sends one query with all samples to the SUT at the beginning of the execution. 
According to Model-equivalence rules, dynamically switching between one or more batch sizes within the scenario's limits is allowed. Following this rule, we tweak offline batch size for a given SUT in order to prevent device out of memory, as well as maximize inference throughput. 
Table~\ref{tab:mlperf} shows specific metrics measured in each scenario to evaluate SUT performance. For single-stream scenario, $90\%$-ile measured latency are measured so that $90\%$ of total queries would be done in a given time. Similar to multi-stream scenario, but $99\%$ of total queries would be done in a given time. Offline measures average throughput during inferencing in terms of samples per second.

\input{text/mlperf_table.tex}

%% file: text/mlperf_table.tex
\begin{table}[t]
\centering
    \renewcommand{\tabcolsep}{3pt}
\caption{
Criteria of MLPerf Testing Scenario
}

\label{tab:mlperf}
\begin{tabular}{@{} *5l}    
\toprule
{\bf Senario}  &    {\bf Duration} & {\bf Samples/Query}    &  {\bf Performance Metric}    \\
\midrule
\midrule
Single-stream   & \makecell[l]{1024 queries\\and 60 seconds}   & 1       & \makecell[l]{90\%-ile \\measured latency\\ (millisecond)}   \\ \hline
Multi-stream    & \makecell[l]{270,336 queries\\and 600 seconds}     &8             &\makecell[l]{99\%-ile \\measured latency\\ (millisecond)}  \\ \hline
Offline	     & \makecell[l]{1 query\\and 60 seconds}  &At least 24,576     & \makecell[l]{Measured throughput\\(samples/sec)}    \\

\bottomrule
\end{tabular}
\end{table}

%% file: text/quantization.tex
\section{Proposed Approaches and Guidelines for Deep Neural Network Quantization}
\label{sec:quant}

This section provides an overview of relevant quantization concepts and how we used them to generate quantized models for different DL frameworks, including PyTorch, TFLite, ONNX, and OpenVINO.

\subsection{Quantization Methodology}
Most DNN training and inference frameworks use FP32 datatypes by default. However, the weights and activations of DNNs may not require the full range and accuracy of FP32. This provides an opportunity to exploit leaner number formats like FP16, INT16, and INT8 via model quantization. Using smaller datatypes to represent a model can lead to reduced memory footprint, smaller latency, and improved throughput. This approach is especially beneficial for edge devices with limited memory and compute resources. There are several technicalities involved when it comes to mapping the full range of FP32 values into a smaller representation:
\subsubsection{Scaling Factor} In order to convert FP32 values to smaller representations, the scaling factor is used to divide the floating-point values and round them to the nearest integer. We then multiply the output by the scaling factor again. The scaling factor is critical for minimizing the difference between the original and quantized values, which in turn minimizes the quantization error.
\subsubsection{Clipping Range} The clipping range determines the range of values that will be retained after quantization. All other values that fall outside this range will be clipped to the minimum or maximum bounds of this range. Clipping is performed to avoid overflow errors in the new representation and to reduce the impact of outliers that can cause issues during the quantization process. The process of choosing the clipping range is called {\em calibration}.
\subsubsection{Quantization Symmetry} Quantization can be either symmetric or asymmetric depending on how we select the clipping range. If the minimum and maximum bounds are set to have the same distance from the central value (usually zero), then the quantized values will be symmetrically distributed. For example, in 8-bit quantization, the clipping range can be between -128 to +127 for symmetric quantization. On the other hand, in asymmetric quantization, the minimum and maximum bounds of the clipping range may have different distances from the center. This results in asymmetric distribution of quantized values. An example for 8-bit asymmetric quantization is to select the clipping range between 0 to 255.
\subsubsection{Static vs. Dynamic Quantization} 
Another important aspect of quantization is the timing of when the scaling factor and clipping range are determined. In static quantization, the quantization parameters are determined, pre-calculated, and fixed during the inference process. Static quantization is often only applied to the weights. In dynamic quantization, on the other hand, the quantization parameters adapt to the input data while the inference is being performed. Dynamic quantization is applied on both the activations and weights and is useful when the data fed to the network varies greatly between different samples. Dynamic quantization is generally considered to be more accurate than static quantization, but it is relatively more compute-heavy compared to static quantization.
\subsubsection{Post-Training Quantization (PTQ) vs. Quantization-Aware Training (QAT)} In post-training quantization (PTQ), we perform quantization on a pre-trained DNN. Weights and activations are determined without retraining the DNN model. PTQ is useful when the data is limited or unlabeled. In contrast, quantization-aware training (QAT) is incorporated into the training process, which requires dataset access. 
During QAT, the network is trained with quantized weights and activations, which usually results in better accuracy at the cost of being a slower process compared to PTQ~\cite{Survey_of_quantization}. Also, this method additively processes pruning to optimize the network. 
In this paper, we focus on the quantization method only. Thus, we narrowed our experiments to the PTQ approach alone.

\input{text/qmethod_table.tex}
\begin{table}[btp]
\centering
    \renewcommand{\tabcolsep}{3pt}
\caption{
Combination of Quantization Methods and DNN Frameworks Used for Performance Evaluation. 
}

\label{tab:quant-methods}
\begin{tabular}{@{} *5l}    
\toprule
{\bf Quantization method}        & {\bf PyTorch}    &  {\bf TFLite}      & {\bf ONNX}     &  {\bf OpenVINO}    \\
\midrule
Default       & \checkmark       & \checkmark         & \checkmark      & \checkmark  \\
INT8-DQ    &                  & \checkmark         & \checkmark    &	\\
INT8-SQ	    & \checkmark       & \checkmark         & \checkmark     &  \\
FP16       &                  & \checkmark         &                &  \\
INT8-OM                      &                  &                    &                & \checkmark \\

\bottomrule
\end{tabular}
\end{table}

The quantization methods that we use in this work are detailed in Table~\ref{tab:quant-methods-num-format}, which include 1) INT8 dynamic asymmetric quantization (INT8-DQ), 2) INT8 static asymmetric quantization (INT8-SQ), 3) half-precision static symmetric quantization (FP16), and 4) 8-bit static symmetric quantization on weights and asymmetric quantization on activations (INT8-OM). Table~\ref{tab:quant-methods} shows the quantization method support offered by different DNN frameworks for Convolutional Neural Networks (CNNs) quantization.

\subsection{Quantization Approachs Based on DL Frameworks}

To evaluate the aforementioned quantization methods and inference scenarios, we select three representative Convolutional Neural Networks (CNNs): DenseNet-12, MobileNetV2, and VGG-19. Depending on the DL framework and quantization technique detailed Tables~\ref{tab:quant-methods-num-format} and~\ref{tab:quant-methods}, we quantize these three models using different configurations to evaluate their performance in terms of latency, throughput, and accuracy. Below are the proposed approaches and guidelines to quantize these models using PyTorch, TFLite, ONNX, and OpenVINO.

\subsubsection{PyTorch Models} The PyTorch versions of DenseNet-121, MobileNetV2, and VGG-19 models and their weights are obtained from TorchVision~\cite{torchvision}. PyTorch's API provides two different quantization methods called Eager Mode Quantization and FX Graph Mode Quantization. Eager Mode Quantization is an experimental feature. The user needs to perform manual operator fusion for quantization and dequantization. FX Graph Mode Quantization is a newly offered feature that automates quantization. In this work, we use the FX Graph Mode Quantization method to perform static quantization over the default FP32 model to obtain the INT8-SQ models. We only perform static quantization using PyTorch due to the framework's lack of support for dynamic quantization over convolution layers.

\subsubsection{TFLite Models}
The TFLite versions of DenseNet-121, MobileNetV2, and VGG-19 models and their weights are obtained from Keras Applications\cite{keras_application}. The quantized models were generated using TFLite default quantization converter. There are four quantization variants of TFLite models: 1) Dynamic Quantization (DQ), 2) Static Quantization (SQ), 3) FP16 quantization (FP16), 4) 16-bit activations with 8-bit weights (Mixed). Dynamic quantization is the default setting of the TFLite converter. The ``dynamic-range'' operators dynamically quantize activations based on their range to 8-bits and perform computations with 8-bit weights and activations. Compared to full fixed-point static quantization, the outputs of the dynamic-range operators are stored in floating-points, resulting in lesser speedups for the dynamic quantized method when compared to the full fixed-point one. 
Static quantization, as known as full integer quantization in TFLite, offers additional latency enhancements, decreases in peak memory usage, and improved compatibility with hardware devices that only support integers. We implemented a representative dataset feeder using the ImageNet 2012 calibration dataset, which is provided by the MLPerf Inference Benchmarks. By using this representative dataset, calibration was performed on the SQ models.

\subsubsection{ONNX Models}
The ONNX versions of MobileNetV2 and VGG19 model were obtained from ONNX Model Zoo\cite{onnx_model_zoo}. The ONNX version of DenseNet-121 model was obtained through the export of the TorchVision version of DenseNet-121 from PyTorch. Quantized ONNX models can be represented in either operator-oriented (QOperator) or tensor-oriented (QDQ; Quantize and DeQuantize) methods. In the operator-oriented representation, all quantized operators have their own ONNX definitions. In contrast, in the tensor-oriented representation, quantization and dequantization functions are inserted between the original operators. The operator-oriented representation can be converted to its equivalent QDQ format\cite{ONNX_Runtime}. In our evaluation, the ONNX Runtime APIs were used to perform dynamic and static quantization over the original ONNX format model.

\subsubsection{OpenVINO Models}
The OpenVINO framework supports both the ONNX format and the OpenVINO Intermediate Representation (IR) format. However, the IR format is recommended as it allows for more optimizations when using the OpenVINO Model Optimizer (MO), which only supports the IR format. 
To obtain quantized IR models, we first convert the original DenseNet-121, MobileNetV2, and VGG-19 ONNX models to FP32 IR models using the MO with default settings. 
Then, using OpenVINO Post-training Optimization Tool (POT), we perform uniform integer quantization on the obtained IR models. We implement a calibration dataset feeder using the same ImageNet 2012 calibration dataset provided by MLPerf, which provides samples needed for calibration. 

The OpenVINO Post-training Optimization Tool (POT) offers a range of hyperparameters to fine-tune the quantization algorithms, giving users flexibility in choosing the number of quantized bits, number of calibration samples, symmetric/asymmetric quantization, granularity, range estimators, etc. To further improve quantization quality, we tune POT hyperparameters in five separate ways and pick one hyperparameter set with the best balance between accuracy, performance, and model size. Using this hyperparameter set, we conduct quantization on the non-quantized IR models.

%% file: text/qmethod_table.tex
\begin{table}[btp]
\centering
    \renewcommand{\tabcolsep}{3pt}
\caption{The Analyzed Quantization Methods. 
}

\label{tab:quant-methods-num-format}
\begin{tabular}{@{} *6l}    
\toprule
\makecell{\bf Quantization\\ \bf method}      &  \makecell{\bf Dynamic/ \\ \bf Static}  & \makecell{\bf Bits}  & {\bf Data Type}    &  {\bf Symmetric/Asymmetric}    \\
\midrule
\midrule

Default      & N/A                &  32                & FP32         & N/A  \\  \hline
INT8-DQ      & Dynamic            &  8                  & INT8       & Asymmetric  \\   \hline
INT8-SQ	    & Static              &  8               & INT8       & Asymmetric  \\  \hline
FP16       & Static            &  16                  & FP16       & Symmetric  \\     \hline
INT8-OM        & Static            &  8              & INT8      & \makecell[l]{Symmetric on weights,\\ Asymmetric on activations}  \\

\bottomrule
\end{tabular}
\end{table}

%% file: text/setup.tex
\section{Experimental setup}
\label{sec:setup}

This section details the hardware platform used for conducting this study. We also enumerate the state-of-the-art models and DL frameworks used along with models and datasets. Details on the selected quantization methods are also presented.

\subsection{Hardware Configurations}

The hardware configurations used in this paper are presented in Table~\ref{tab:specs}. 
We rely on two HPC platforms---TACC Frontera and an internal system at The Ohio State University called RI2---as well 
as an edge device---Raspberry Pi 4B---to conduct our characterization study.

\begin{table}[htbp]
\centering
    \renewcommand{\tabcolsep}{3pt}
\caption{
Hardware specification of the Raspberry Pi 4B, in-house RI2 System, and the TACC Frontera System. 
}

\label{tab:specs}
\begin{tabular}{@{} *5l}    
\toprule
{\bf Specification} & {\bf Raspberry Pi 4B} & {\bf Frontera}     &  {\bf Ri2}    \\
\midrule
Processor Family	&	Cortex-A72 (ARMv8)	        &	Xeon Cascade Lake	    &  Xeon Skylake\\
Processor Model	    &	Broadcom BCM2711	        &	Platinum 8280	    & Gold 6132 \\
Clock Speed	        &	1.5 GHz     	            &	2.7 GHz	                & 2.6 GHz\\
Sockets	            &	1	            &	2	            & 2\\
Cores Per socket	&	4	            &	28	            & 14\\
RAM	        &	8 GB	        &	192 GB 	            & 192 GB \\
\bottomrule
\end{tabular}
\end{table}

\subsection{Software Packages and Versions}

MLPerf Edge Inference benchmark suite v2.1 has been used in this study.
This suite contains the LoadGen python module---responsible for generating input traffic----that is built with the default setting.

The 2.9.1 version of TensorFlow Lite module in Intel-Tensorflow~\cite{intel_tensorflow} package is used on Frontera and RI2 systems. Also, the 2.9.1 version of tflite-runtime is utilized on the Raspberry Pi 4B device. The 1.12.1 version is used with PyTorch and ONNX runtime on all platforms. OpenVINO version 2022.2.0 is employed for Frontera and RI2 systems. OpenVINO is built from source code for Frontera and RI2 systems following the official build guide for CentOS. We did not use the model optimization features on frameworks not to impact the quantization characteristic.

\subsection{Models and Datasets}    

In this study, we used three representative popular image classification DNN models, DenseNet-121, MobileNetV2, and VGG-19, are used: 
\begin{itemize}
   \item Dense Convolutional Network (DenseNet) has a feed-forward fashion between layer-to-layer connections. It embraced the observation that convolutional networks can be substantially deeper, more accurate, and more efficient to train if they contain shorter connections between layers close to the input and those close to the output.
   \item MobileNet is a class of efficient models for mobile and embedded vision applications. It is based on a streamlined architecture that uses depth-wise separable convolutions to build light weight deep neural networks.
   \item VGG is a class of deep convolutional networks which use architecture with tiny (3x3) convolution filters. By increasing the depth to 16-19 weight layers. It significantly improved accuracy in the large-scale image recognition setting compared with its prior state-of-the-art results. 
\end{itemize}

The validation dataset of ImageNet Large Scale Visual Recognition Challenge 2012 (ILSVRC2012)~\cite{ILSVRC15} is used to input data for all models under test. 
Input images are resized for the size that is suggested on each model. 
Final values are rescaled to $[0.0, 1.0]$ and then normalized using the mean value of $[0.485, 0.456, 0.406]$ and standard deviation of $[0.229, 0.224, 0.225]$.

%% file: text/evaluation.tex
\section{Evaluation and analysis}
\label{sec:evaluation}

In this section, we show the results of performance characteristics on quantization and analysis the results. The following quantization techniques were used: FP32 = Default, INT8-SQ = Static Quantization with INT8 Format, INT8-DQ = Dynamic Quantization with INT8 Format, FP16 = Half-Precision Format, INT8-OM = 8-bit symmetric quantization on weights and asymmetric quantization on activations.

\subsection{Model Accuracy and Size of Quantized Models}
\label{sec:eval:accuracy}

\begin{figure*}[!t]
     \centering
     \begin{subfigure}[b]{.3\textwidth}
         \centering
         \includegraphics[width=\textwidth]{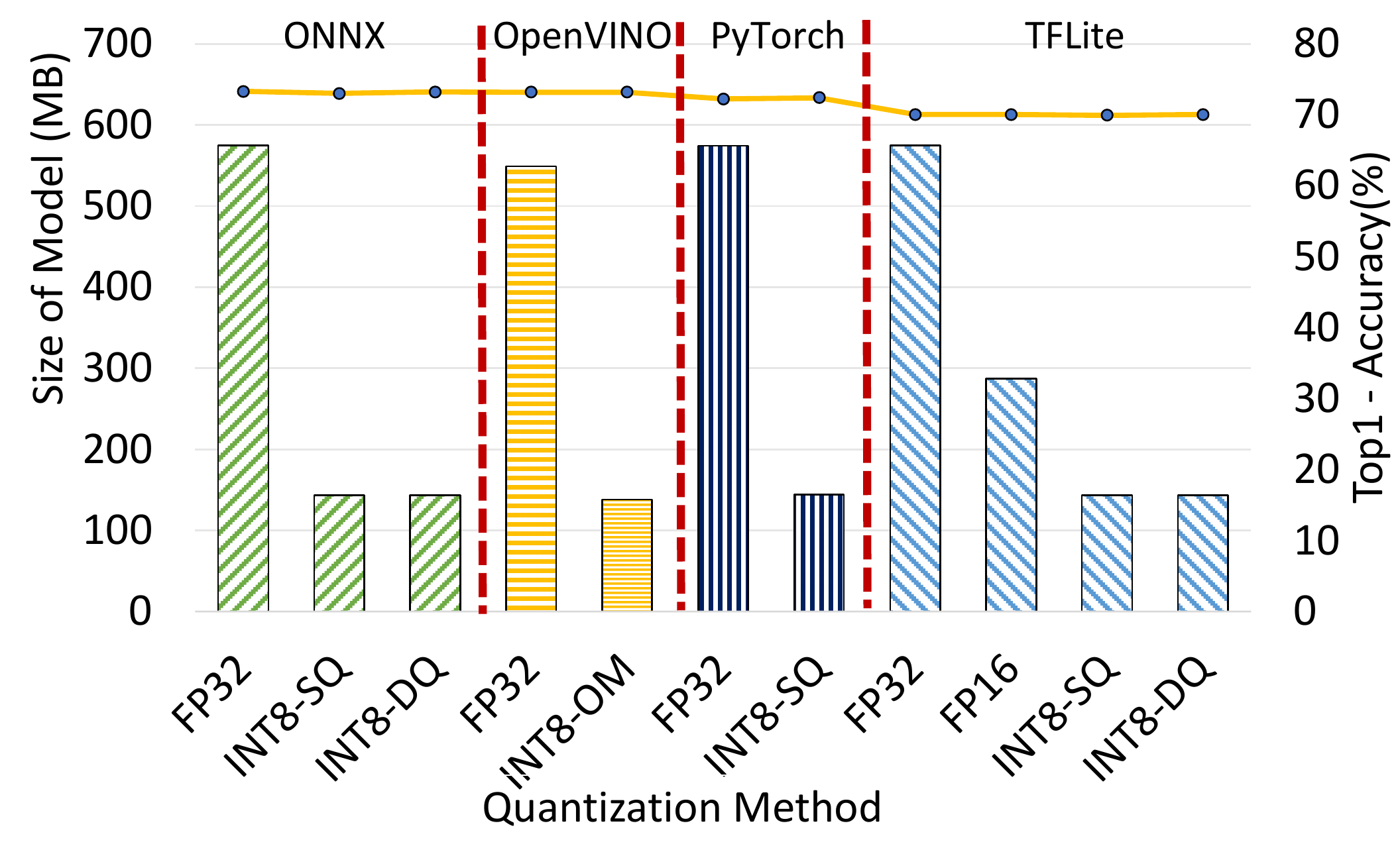}
         \caption{VGG-19}
         \label{fig:accuracy and model size VGG19}
     \end{subfigure}
     \hfill
     \begin{subfigure}[b]{.3\textwidth}
         \centering
         \includegraphics[width=\textwidth]{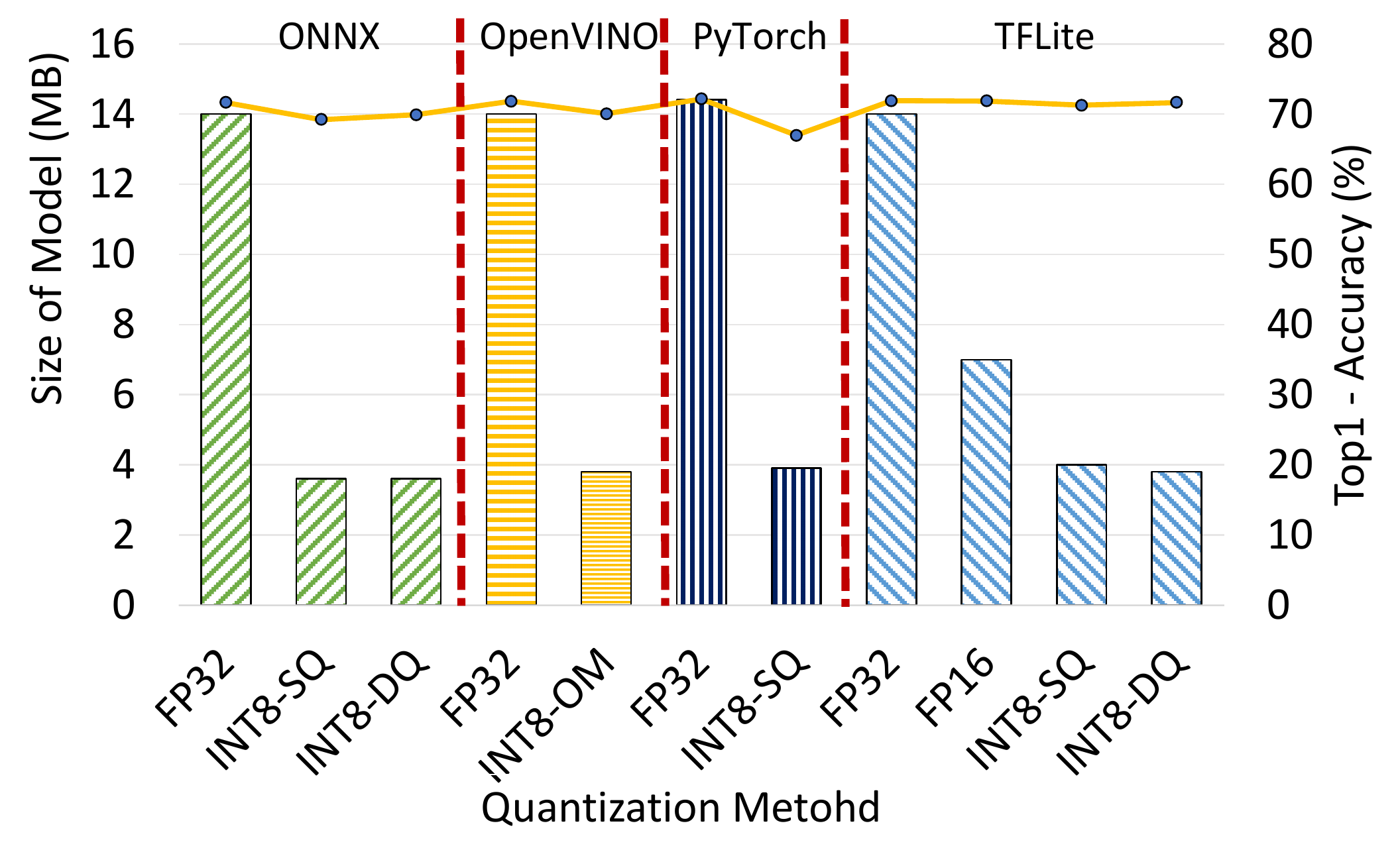}
         \caption{MobileNetV2}
         \label{fig:Accuracy and model size mobilenet}
     \end{subfigure}
     \hfill
     \begin{subfigure}[b]{.3\textwidth}
         \centering
         \includegraphics[width=\textwidth]{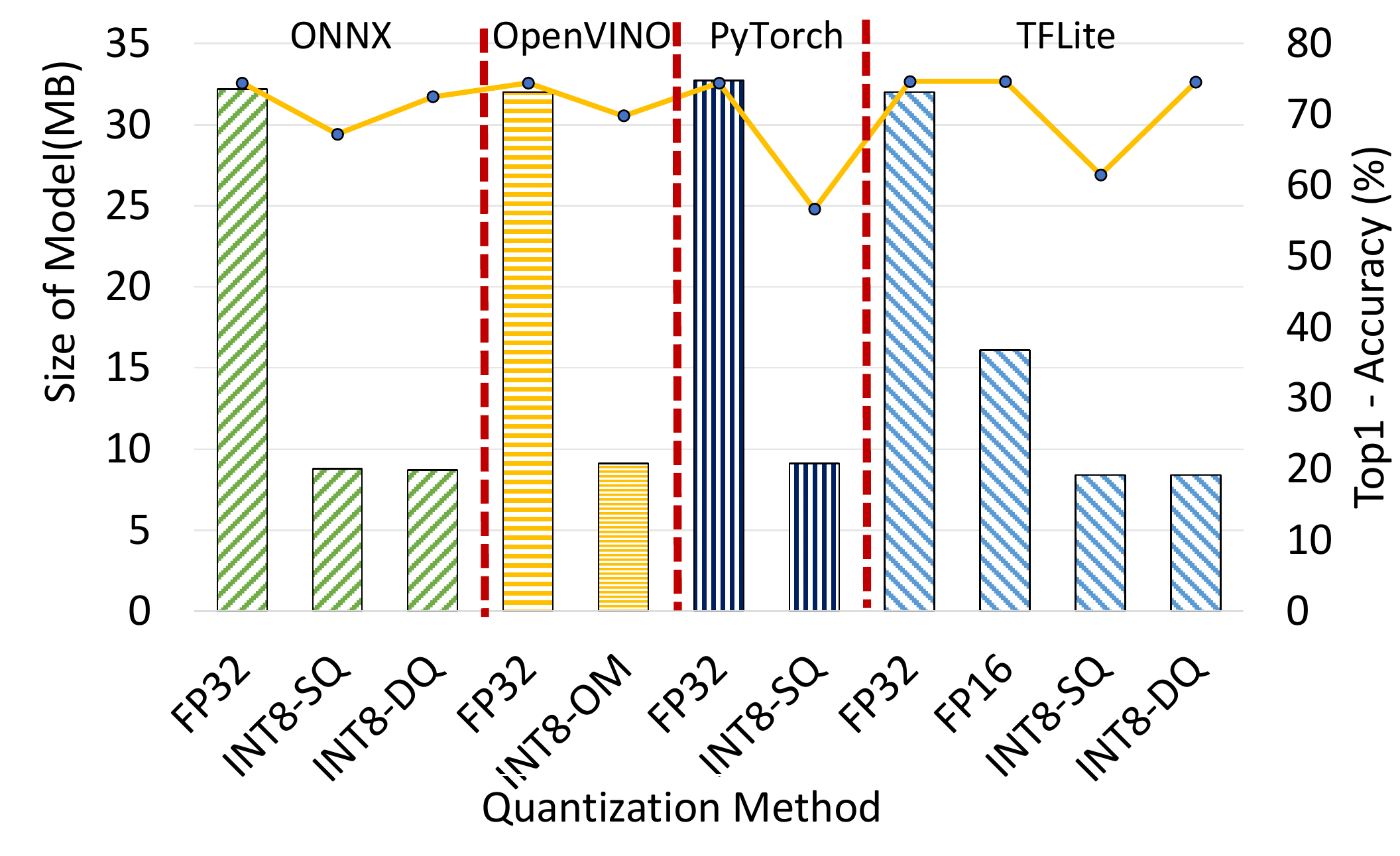}
         \caption{DenseNet-121}
         \label{fig:Accuracy and model size denseNet121}
     \end{subfigure}
        \caption{Model accuracy and size for VGG-19, MobileNetV2, and DenseNet-121 on all four frameworks (ONNX, PyTorch, TFLite, and OpenVINO) using the ImageNet validation dataset. Accuracy is plotted with the line on the {\tt y2} axis}  
        \label{fig:accuracy-and-model-size}
\end{figure*}

Figure~\ref{fig:accuracy-and-model-size} shows the overall experimental results of the model accuracy and size, including VGG-19, MobileNetV2, and DenseNet-121 on all four frameworks (ONNX, PyTorch, TFLite, and OpenVINO) using the ImageNet validation dataset.
Since INT8-SQ, INT8-DQ, and INT8-OM utilize the 8-bit integer representation, we witness the model size reduction by a quarter to the original FP32 model. The model size of the FP16 variant is reduced by half. We note that while the model sizes are reduced substantially, the accuracy of quantized models is as good as the original FP32 models. The only exception is the INT8-SQ variants because of the use of static clipping range during the model calibration. The drop in accuracy with the INT8-SQ quantized model is the most visible for PyTorch and TFLite for the DenseNet-121 model.

\subsection{Evaluating Inference Latency and Throughput using MLPerf Benchmark}

\begin{figure*}[!t]
     \centering
     \begin{subfigure}[b]{.3\textwidth}
         \centering
         \includegraphics[width=\textwidth]{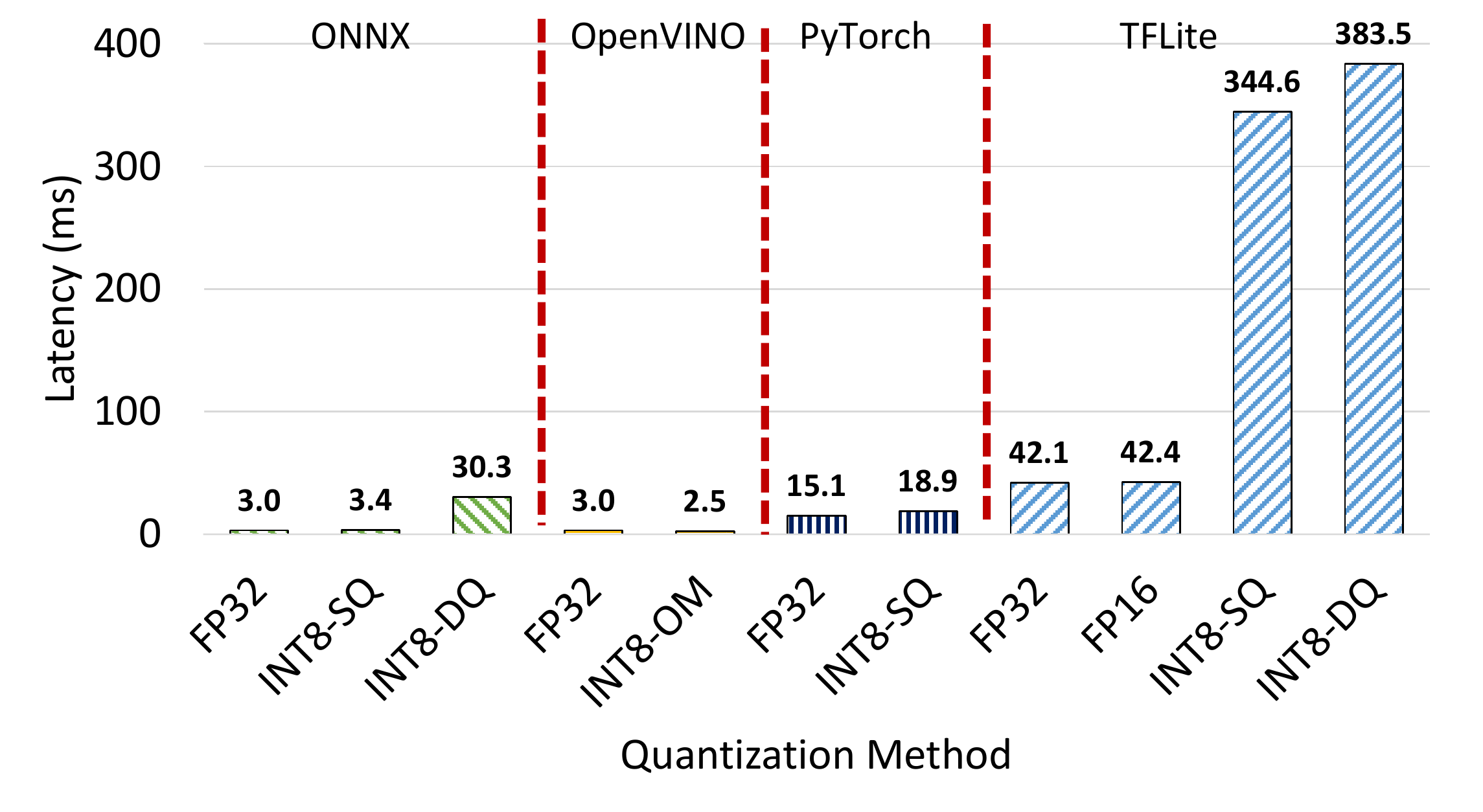}
         \caption{Single-stream}
     \end{subfigure}
     \hfill
     \begin{subfigure}[b]{.3\textwidth}
         \centering
         \includegraphics[width=\textwidth]{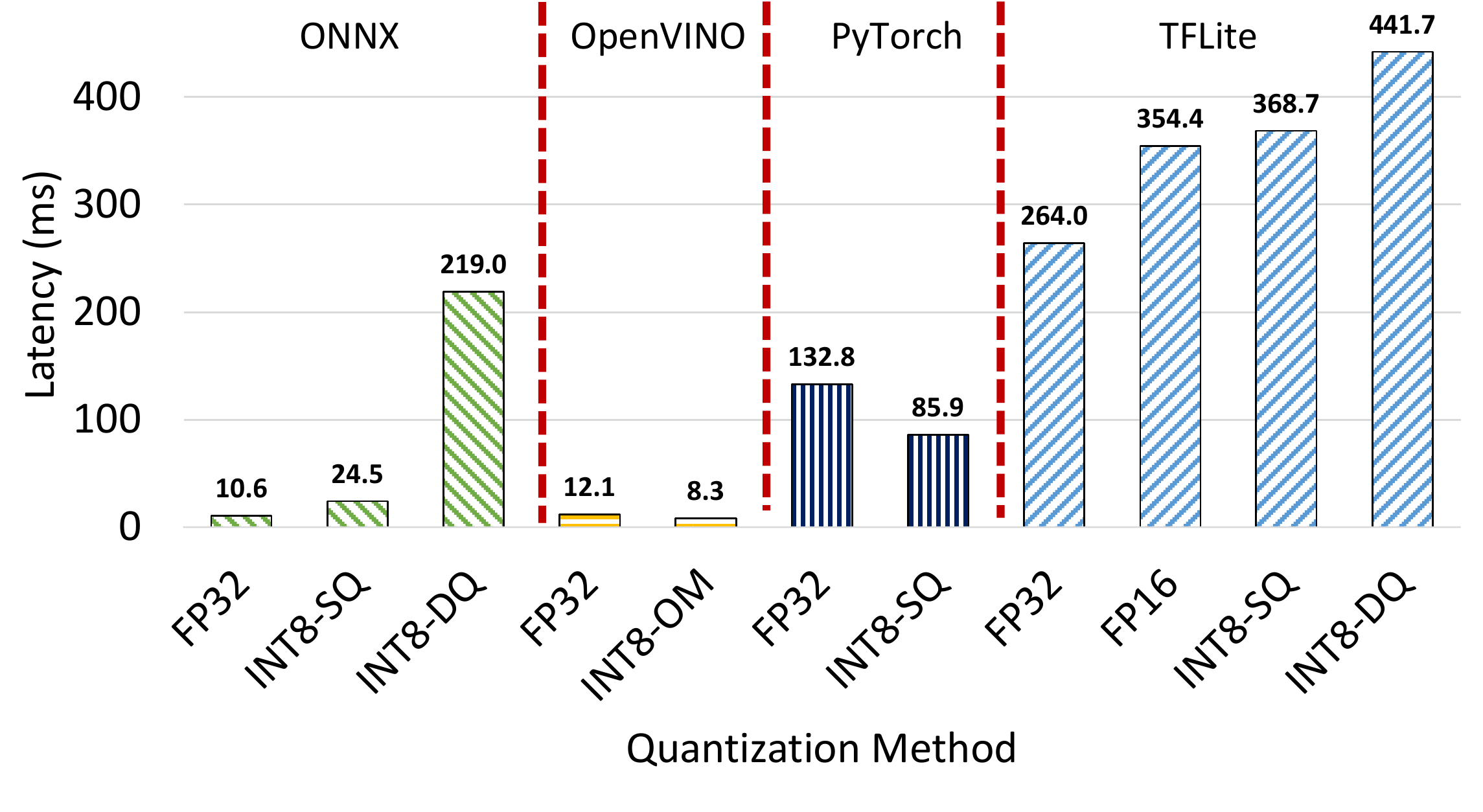}
         \caption{Multi-stream}
     \end{subfigure}
     \hfill
     \begin{subfigure}[b]{0.3\textwidth}
         \centering
         \includegraphics[width=\textwidth]{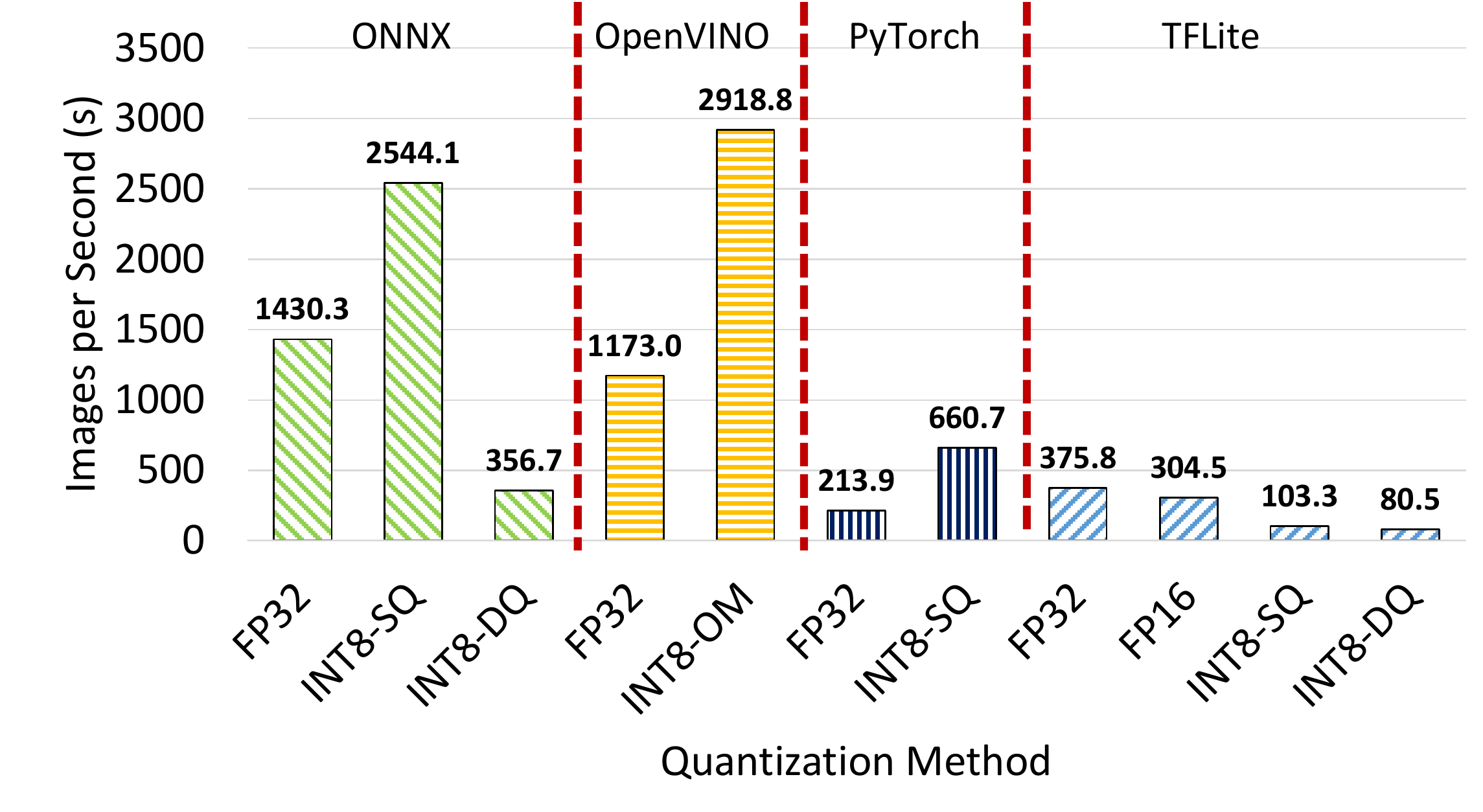}
         \caption{Offline}
     \end{subfigure}
        \caption{Inference performance of ONNX, OpenVINO. PyTorch, and TFLite using MLPerf Edge benchmarks with single-stream, multi-stream, and offline scenarios on the TACC Frontera System. The model is MobileNetV2.}
        \label{fig:Frontera_mobilenet_all_senario_and_all_frameworks}
\end{figure*}

\begin{figure*}[!htbp]
     \centering
     \begin{subfigure}[b]{.3\textwidth}
         \centering
         \includegraphics[width=\textwidth]{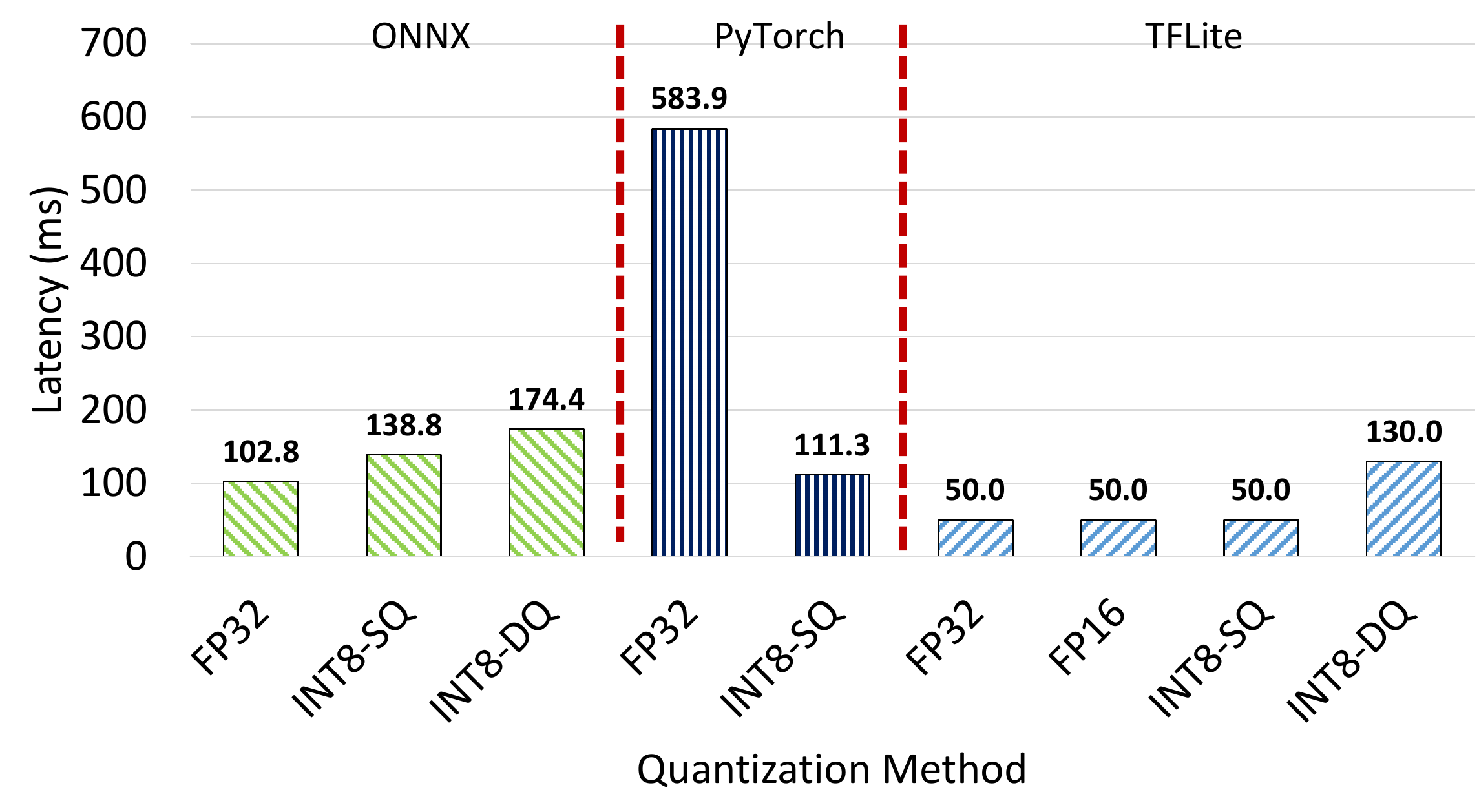}
         \caption{Single-stream}
     \end{subfigure}
     \hfill
     \begin{subfigure}[b]{.3\textwidth}
         \centering
         \includegraphics[width=\textwidth]{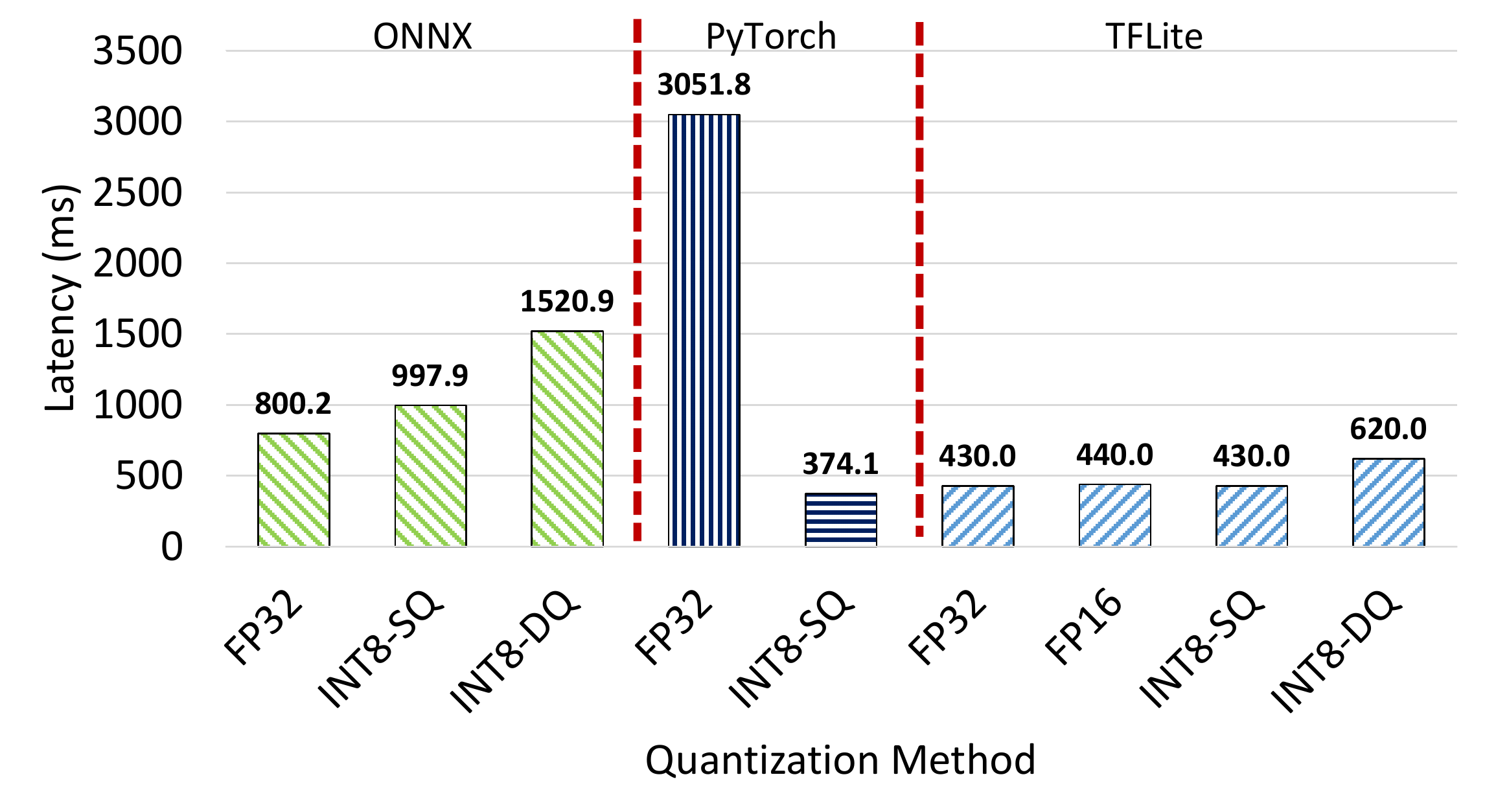}
         \caption{Multi-stream}
     \end{subfigure}
     \hfill
     \begin{subfigure}[b]{0.3\textwidth}
         \centering
         \includegraphics[width=\textwidth]{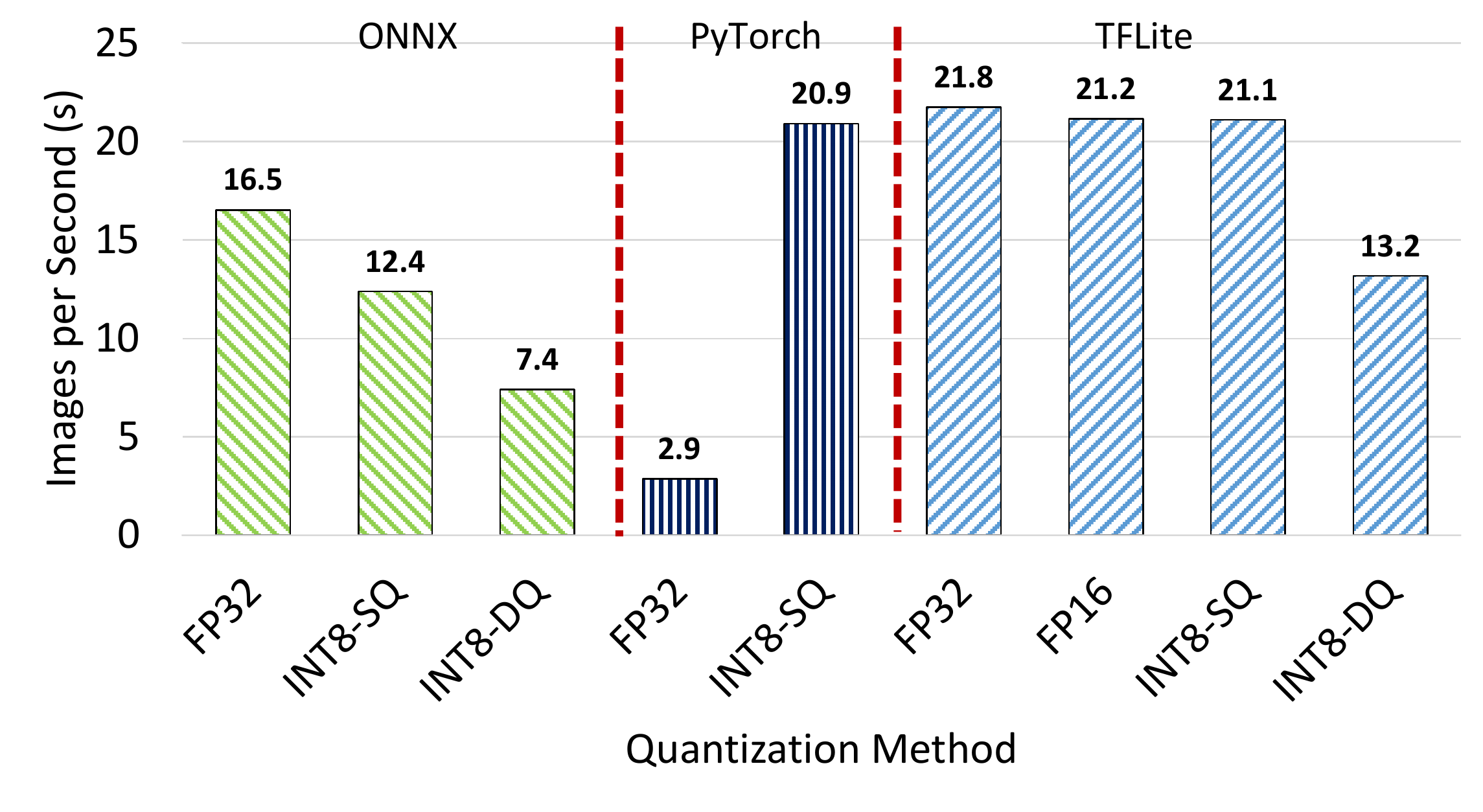}
         \caption{Offline}
     \end{subfigure}
        \caption{Inference performance of ONNX, PyTorch, and TFLite using MLPerf Edge benchmarks with single-stream, multi-stream, and offline scenarios on the Raspberry Pi 4B device. The model is MobileNetV2.}
\label{fig:RPI4_mobilenet_all_senario_and_all_frameworks}
\end{figure*}

\textbf{MobileNetV2.} Figures~\ref{fig:Frontera_mobilenet_all_senario_and_all_frameworks} and~\ref{fig:RPI4_mobilenet_all_senario_and_all_frameworks} present the MLPerf Edge inference benchmarks performance numbers---single-stream, multi-stream, and offline scenarios---for the MobileNetV2 model on the TACC Frontera system and the Raspberry Pi 4B device, respectively. The reason for choosing MobileNetV2 is the small model size and high accuracy (as discussed in Section~\ref{sec:eval:accuracy}).
Figure~\ref{fig:Frontera_mobilenet_all_senario_and_all_frameworks} shows that ONNX and OpenVINO are the most optimized frameworks for Intel CPUs on the Frontera system for the default FP32 format. The OM performance of OpenVINO shows performance benefits over FP32. %
The DQ method---for both Frontera and Raspberry---is always slower than FP32 because DQ exhibits overhead due to scale factor calculation at runtime. PyTorch is slower than ONNX and OpenVINO, but SQ improves the performance as it employs the FBGEMM library, which is optimized for low-precision calculation on the x86 architecture.  
TFLite shows the lowest performance among the frameworks, and quantization does not enhance the latency and throughput. The main reason is TFLite primarily targets ARM and embedded devices and is not optimized for Intel CPUs. Also, we observe that SQ only shows performance benefits in the offline scenario for the ONNX framework. This is because ONNX provides better quantization inferences with mini-match on Intel CPUs. 
Figure~\ref{fig:RPI4_mobilenet_all_senario_and_all_frameworks} presents the performance evaluation on the Raspberry Pi 4B device. Here, we exclude the OpenVINO framework since it mainly targets Intel CPU. Like Frontera, we note that DQ does not improve performance. Contrary to Frontera, where TFLite exhibited the worst performance, TFLite here shows the best performance compared to other frameworks, including ONNX and PyTorch. 
We do not observe any benefits of using quantization with the ONNX framework. On the other hand, the SQ performance for PyTorch is significantly faster than FP32. The main reason is that PyTorch uses the optimized QNNPACK as the backend compute library with quantized solutions.  

\begin{figure*}[!htbp]
     \centering
     \begin{subfigure}[b]{.3\textwidth}
         \centering
         \includegraphics[width=\textwidth]{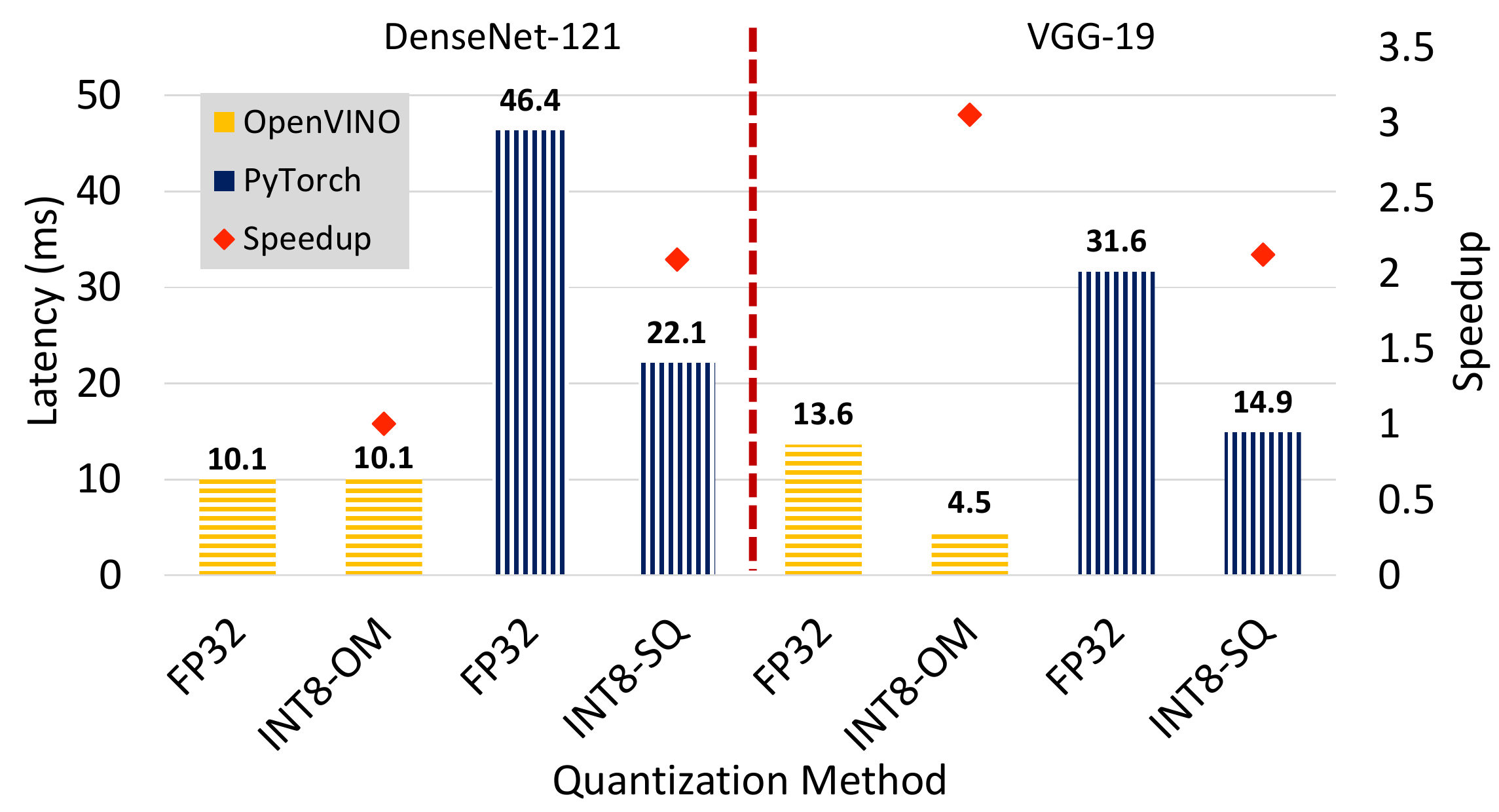}
         \caption{Single-stream}
     \end{subfigure}
     \hfill
     \begin{subfigure}[b]{.3\textwidth}
         \centering
         \includegraphics[width=\textwidth]{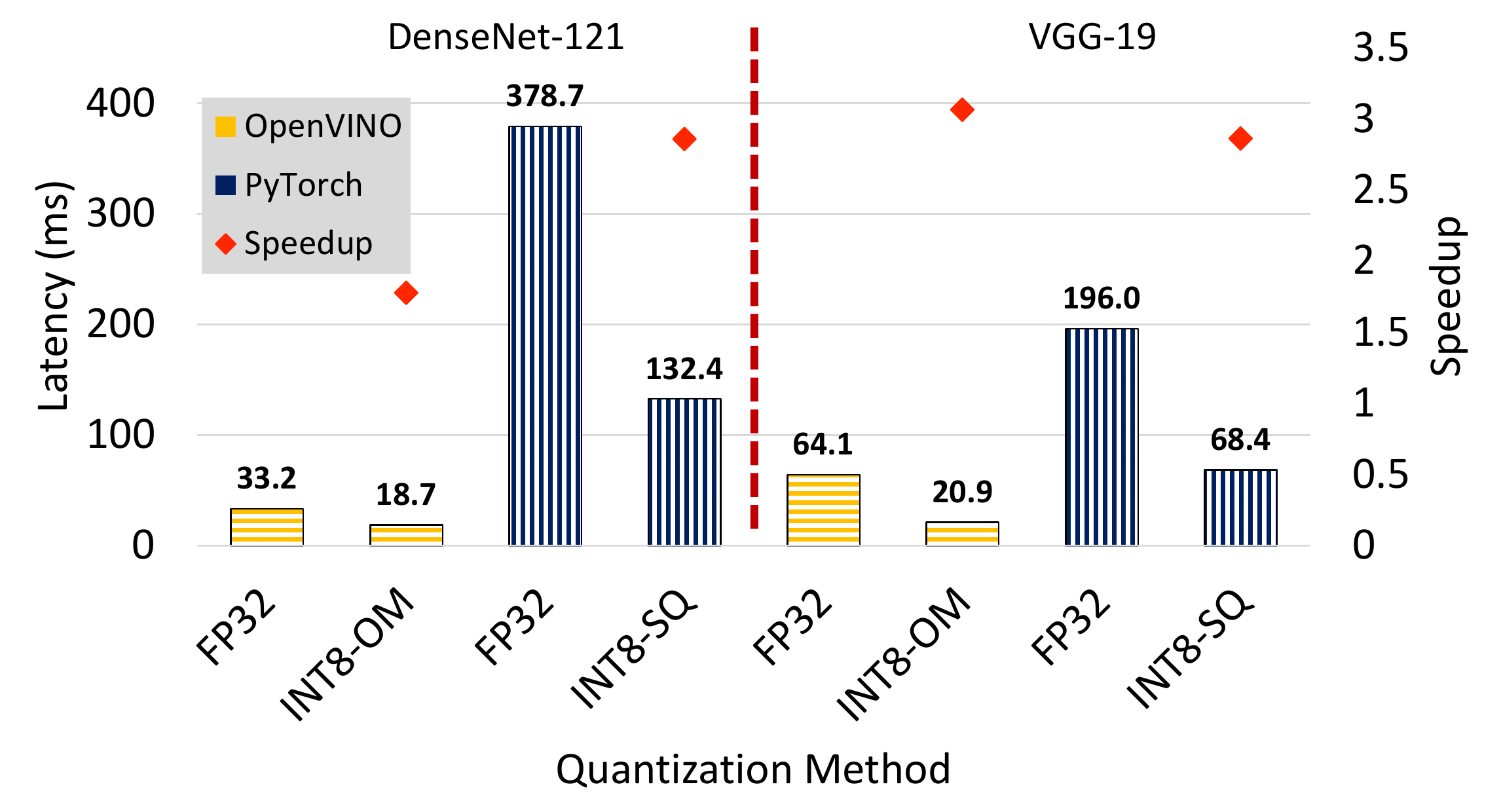}
         \caption{Multi-stream}
     \end{subfigure}
     \hfill
     \begin{subfigure}[b]{0.3\textwidth}
         \centering
         \includegraphics[width=\textwidth]{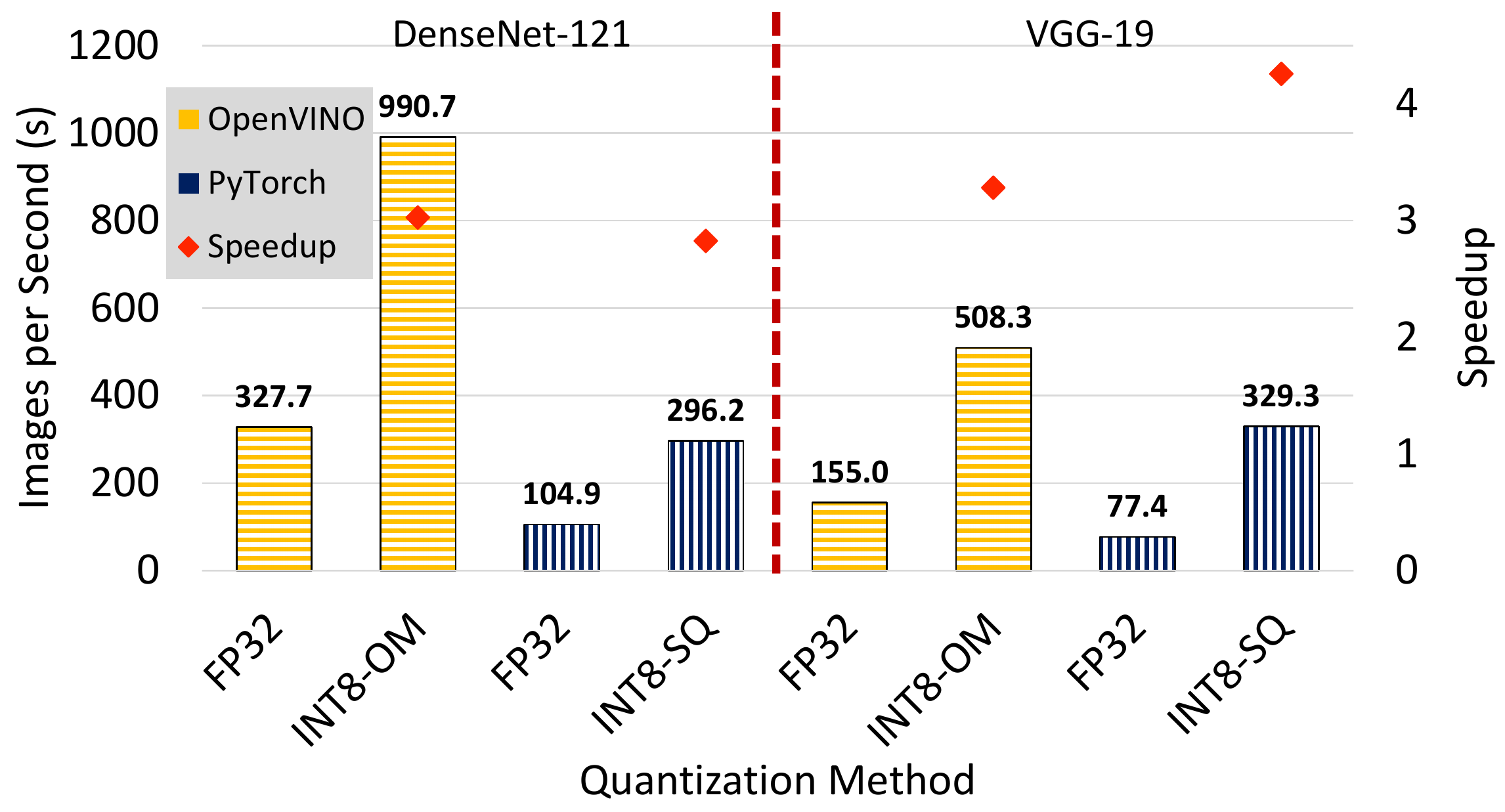}
         \caption{Offline}
     \end{subfigure}
        \caption{Inference performance of OpenVINO and PyTorch using MLPerf Edge benchmarks with single-stream, multi-stream, and offline scenarios on the TACC Frontera System. Models are VGG-19 and DenseNet-121. Speedup is also plotted with diamonds on the {\tt y2} axis using the formula: $\frac{quantized\_performance}{FP32\_performance}$ for offline, $\frac{FP32\_performance}{quantized\_performance}$ for single/multi-stream.}
        \label{fig:VGG19_Densenet121 Frontera speedup}
\end{figure*}

\begin{figure*}[!htbp]
     \centering
     \begin{subfigure}[b]{.3\textwidth}
         \centering
         \includegraphics[width=\textwidth]{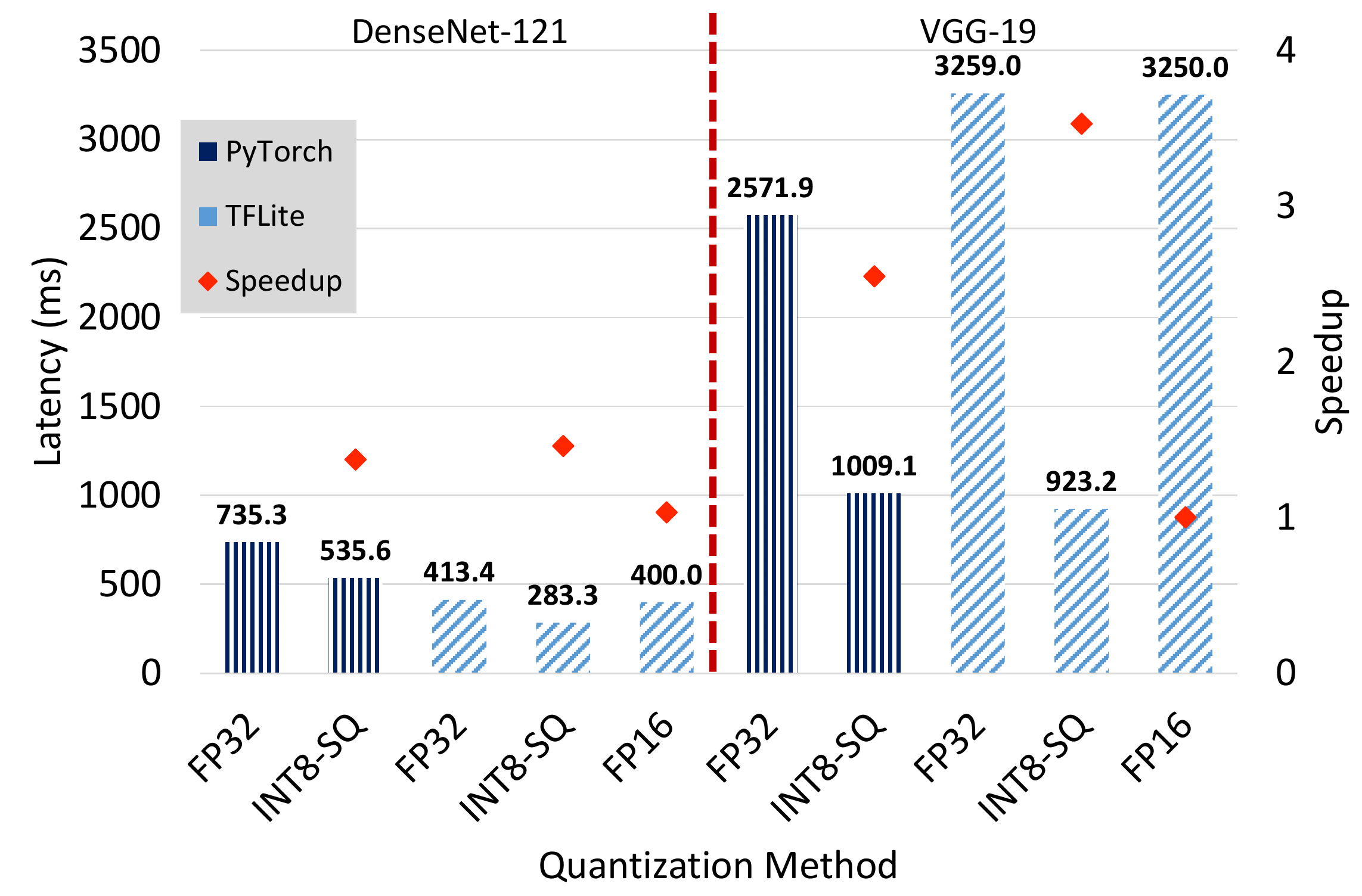}
         \caption{Single-stream}
     \end{subfigure}
     \hfill
     \begin{subfigure}[b]{.3\textwidth}
         \centering
         \includegraphics[width=\textwidth]{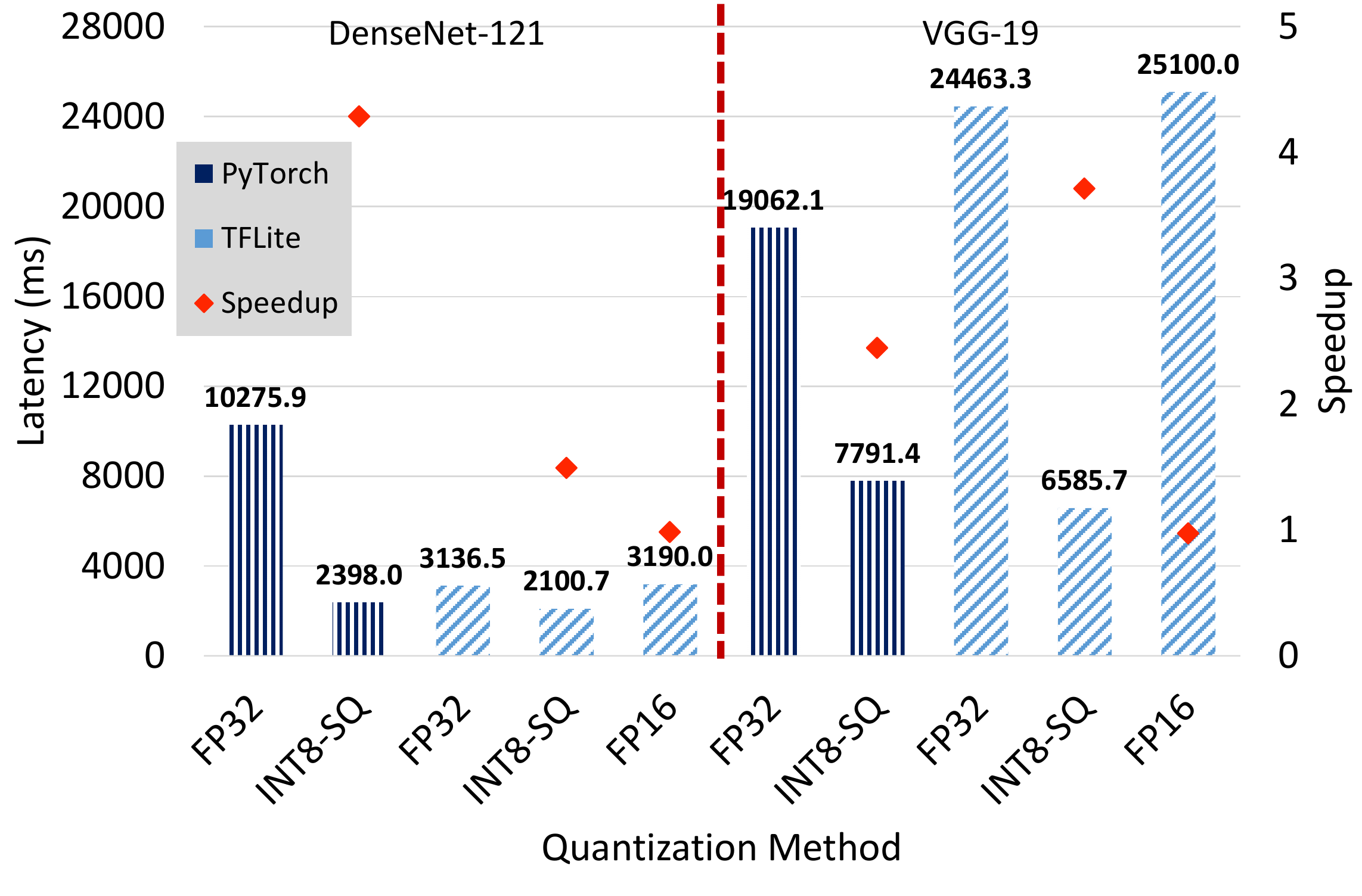}
         \caption{Multi-stream}
     \end{subfigure}
     \hfill
     \begin{subfigure}[b]{0.3\textwidth}
         \centering
         \includegraphics[width=\textwidth]{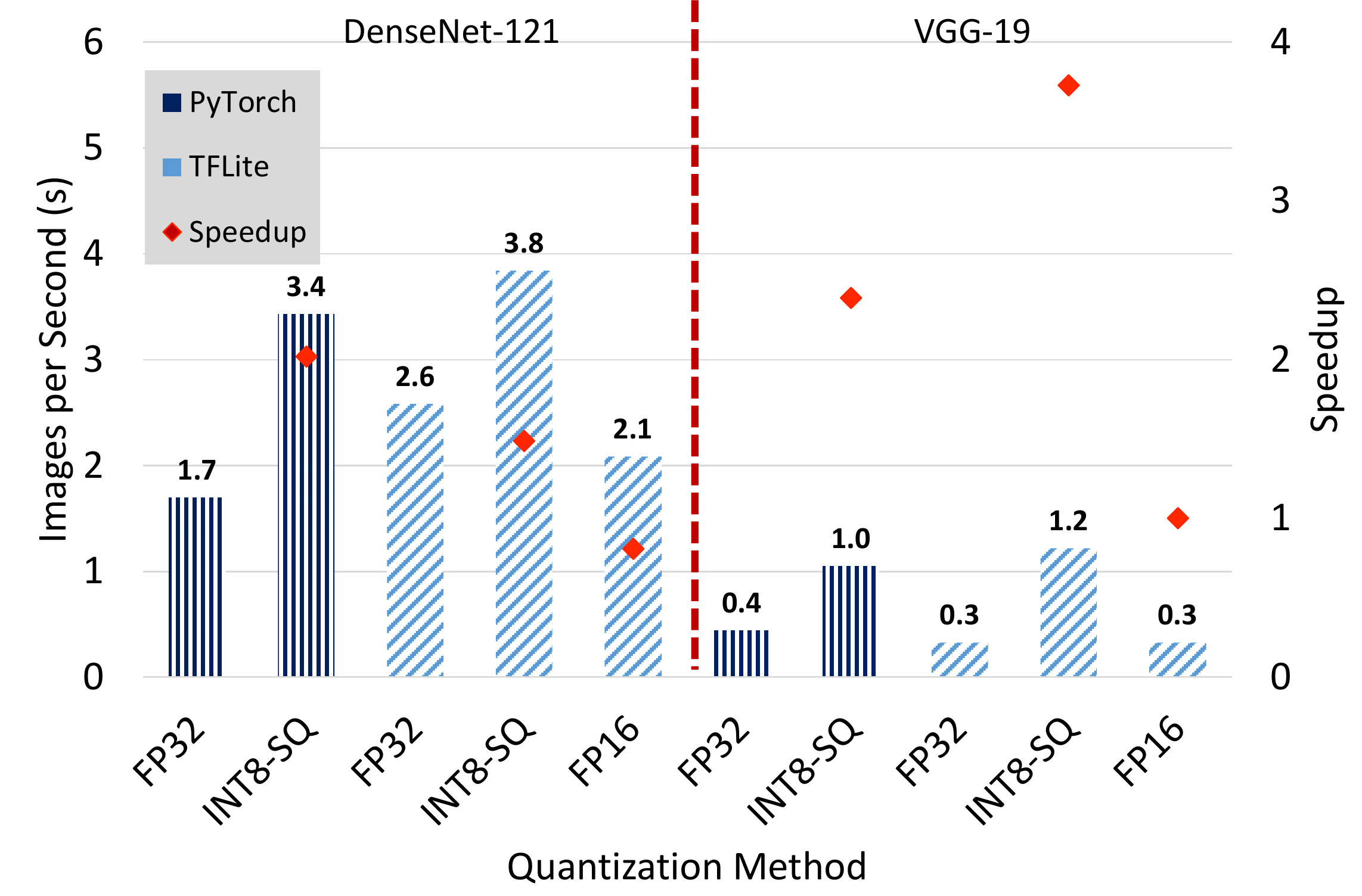}
         \caption{Offline}
     \end{subfigure}
        \caption{Inference performance of TFLite and PyTorch using MLPerf Edge benchmarks with single-stream, multi-stream, and offline scenarios on the Raspberry Pi 4B device. Models are VGG-19 and DenseNet-121. Speedup is also plotted with diamonds on the {\tt y2} axis using the formula: $\frac{quantized\_performance}{FP32\_performance}$ for offline, $\frac{FP32\_performance}{quantized\_performance}$ for single/multi-stream.}
        \label{fig:VGG19_Densenet121 RPI4 speedup}
\end{figure*}

\textbf{DenseNet-121 and VGG-19.}
Figure~\ref{fig:VGG19_Densenet121 Frontera speedup} plots the single-stream, multi-stream, and offline scenario results---with OpenVINO and PyTorch---for DenseNet-121 and VGG-19 on the Frontera system. Results here follow the same trend as discussed earlier for Figure~\ref{fig:Frontera_mobilenet_all_senario_and_all_frameworks}. In addition, we plot the obtained speedup on the {\tt y2} axis that is calculated by the following formula: $ \frac{quantized\_performance}{FP32\_performance}$ for offline scenario and $\frac{FP32\_performance}{quantized\_performance}$ for single/multi-stream scenarios.
Figure~\ref{fig:VGG19_Densenet121 RPI4 speedup} plots the same scenarios/models with TFLite and PyTorch on the Raspberry Pi 4B device.

\subsection{Impact of the Batch Size on the MLPerf Offline Scenario}
\begin{figure*}[!htbp]
     \centering
     \begin{subfigure}[b]{.3\textwidth}
         \centering
         \includegraphics[width=\textwidth]{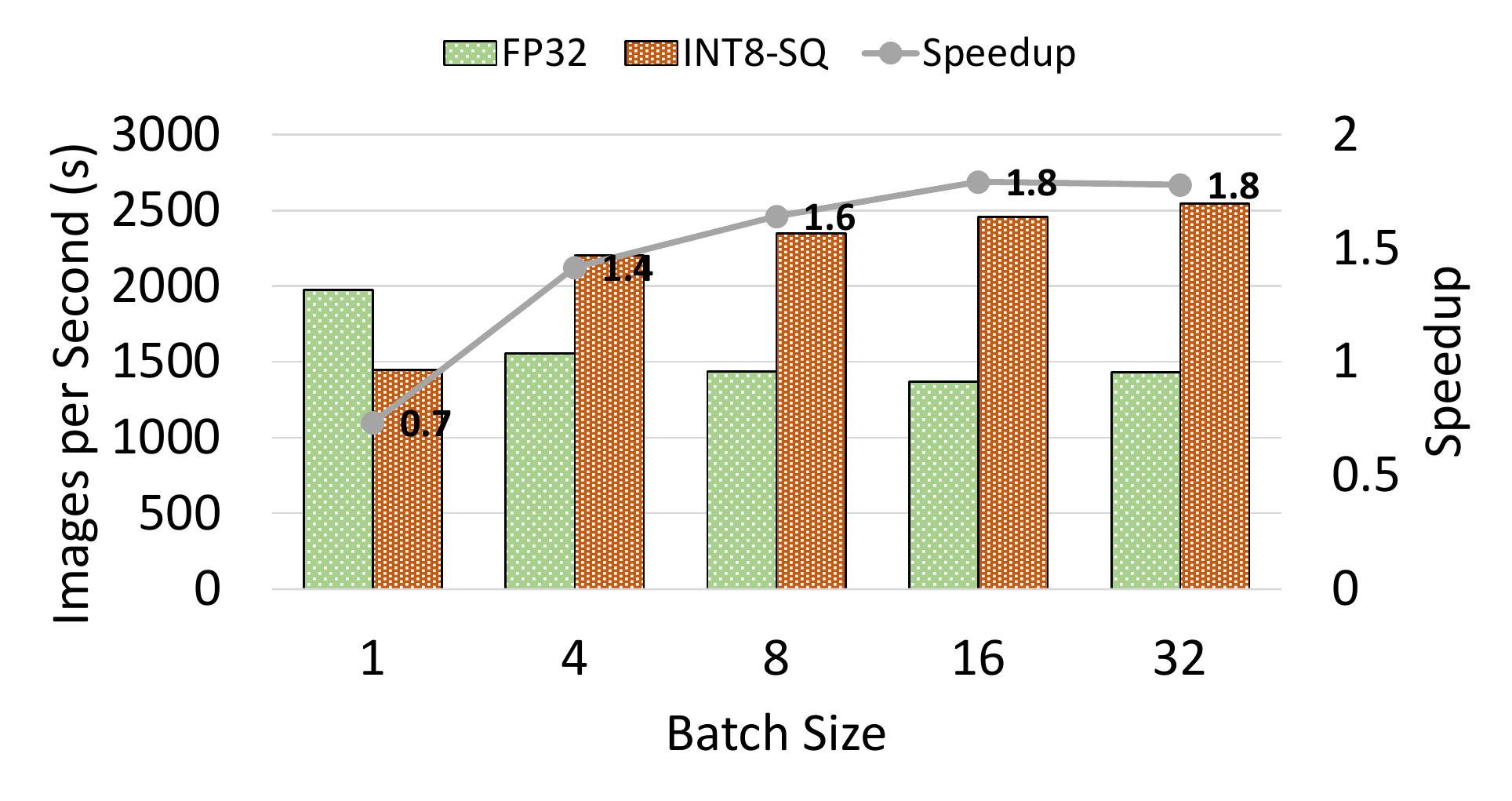}
         \caption{ONNX (Frontera)}
         \label{fig:batch-size-study-frontera-onnx}
     \end{subfigure}
     \hfill
     \begin{subfigure}[b]{.3\textwidth}
         \centering
         \includegraphics[width=\textwidth]{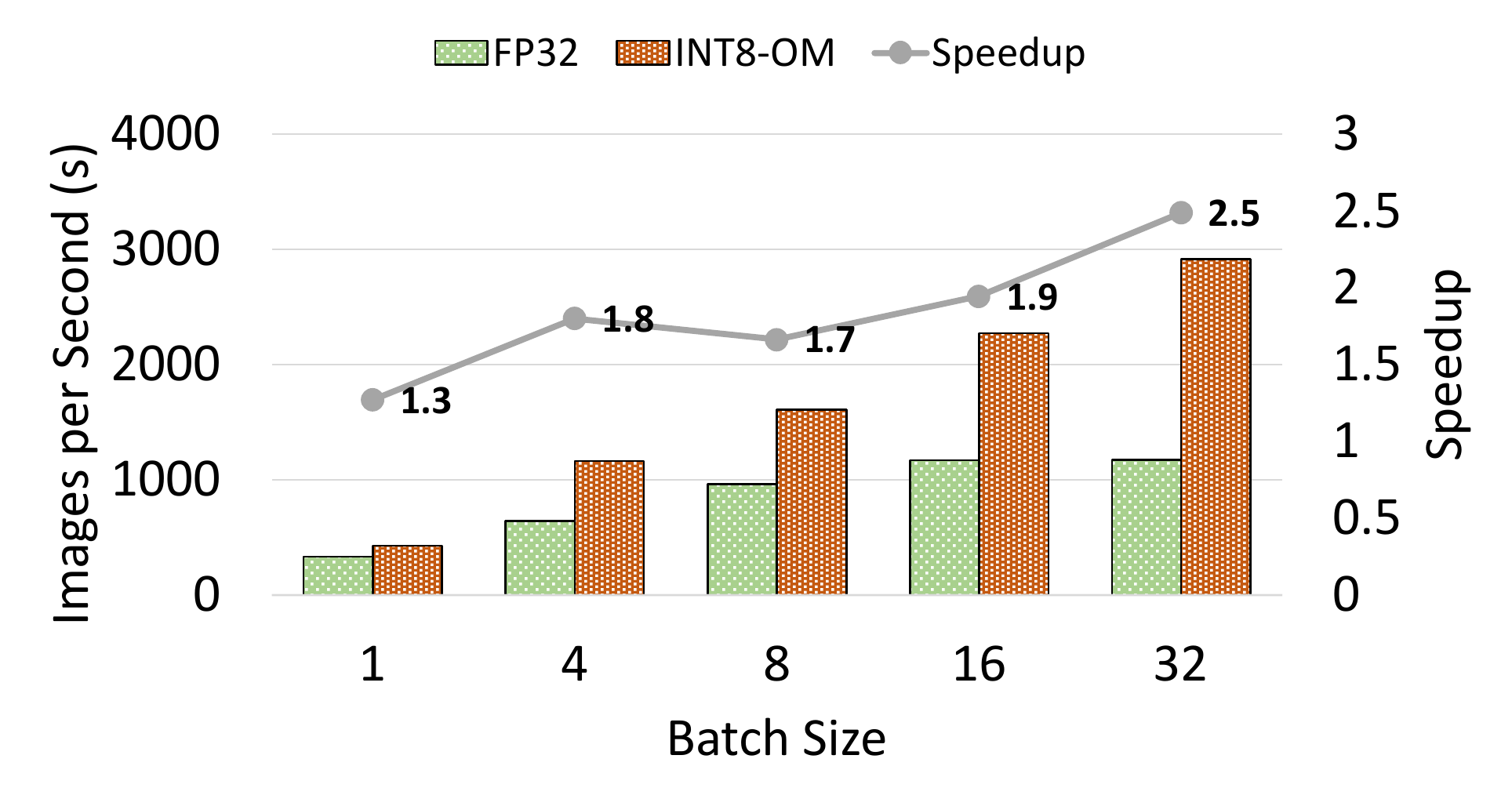}
         \caption{OpenVINO (Frontera)}
         \label{fig:batch-size-study-frontera-openvino}
     \end{subfigure}
     \hfill
     \begin{subfigure}[b]{0.3\textwidth}
         \centering
         \includegraphics[width=\textwidth]{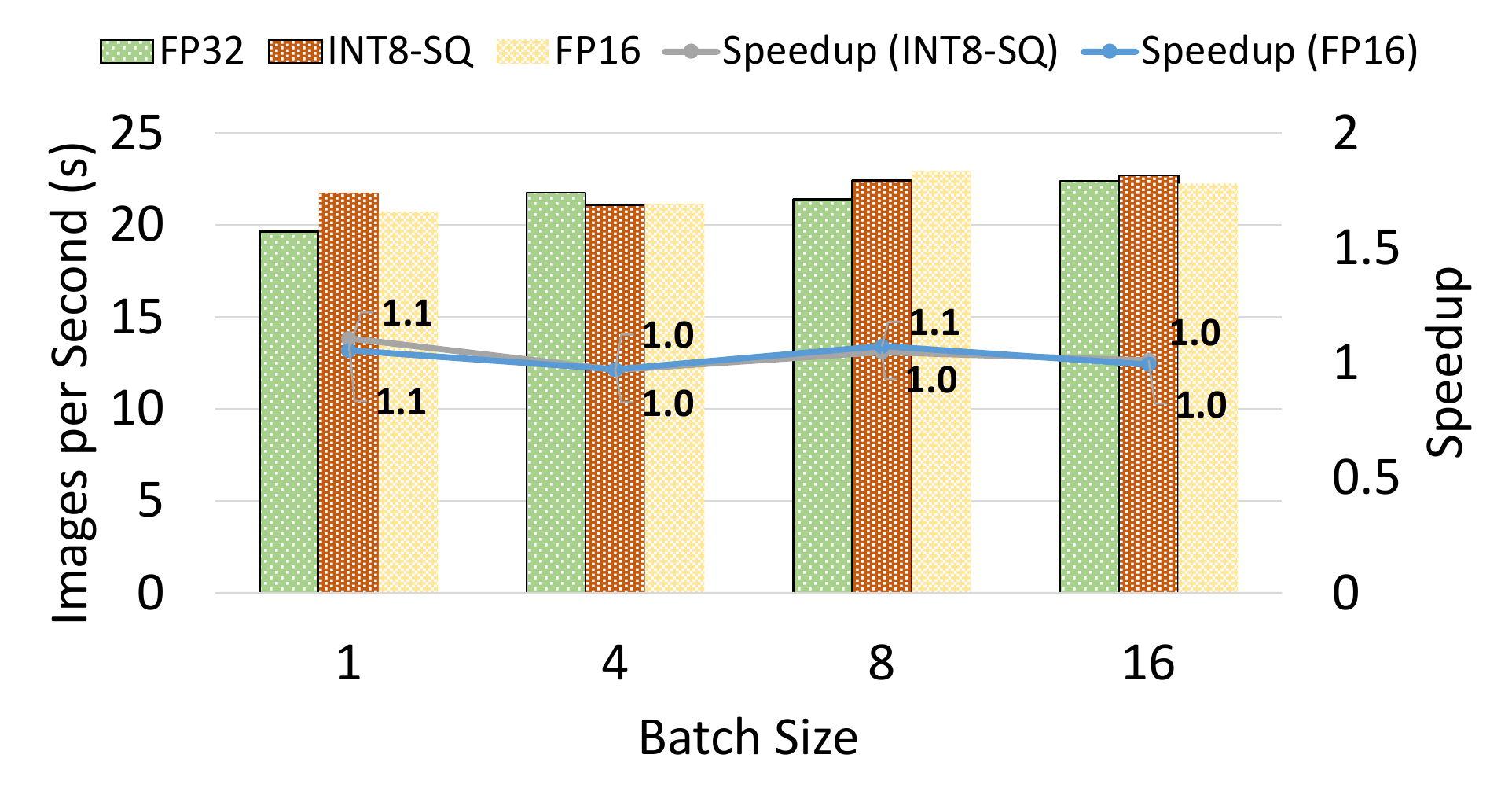}
         \caption{TFlite (Raspberry)}
         \label{fig:batch-size-study-pi4-tflite}
     \end{subfigure}
        \caption{Inference performance of ONNX and OpenVINO (on Frontera) and TFLite (on Raspberry Pi 4B) using MLPerf Edge benchmarks with the offline scenario. The model is MobileNetV2. Speedup is also plotted with a line on the {\tt y2} axis using the formula: $ \frac{quantized\_performance}{FP32\_performance}.$}
        \label{fig:batch-size-study}
\end{figure*}

The batch size hyperparameter controls the number of input images that DNN frameworks can process simultaneously during inference. This sub-section analyzes the impact of batch size using quantized weights/activations.
Figure~\ref{fig:batch-size-study} shows the inference performance---for the offline scenario---of ONNX and OpenVINO (on Frontera) and TFLite (on Raspberry Pi 4B) by varying the batch size from $1$ to $32$. This study is done using the MobileNetV2 model. The speedup is also plotted with a line on the {\tt y2} axis using the formula: $\frac{quantized\_performance}{FP32\_performance}.$
On the Frontera system (Figures~\ref{fig:batch-size-study-frontera-onnx} 
and~\ref{fig:batch-size-study-frontera-openvino}), we observe that the speedup improves by increasing the batch size. The best speedups of $1.8$ and $2.5$ are witnessed for ONNX and OpenVINO, respectively, with $32$ batch size. Also, Figure~\ref{fig:batch-size-study-pi4-tflite} shows that we only witness modest benefits of increasing batch size for the TFLite framework on the Raspberry Pi 4B device. This is because the ARM processor on the device is not able to efficiently process batches of input compared to scalar input.

\subsection{Benefits of Hardware Support for Inference Tasks}

\begin{figure*}[!htbp]
     \centering
     \begin{subfigure}[b]{.3\textwidth}
         \centering
         \includegraphics[width=\textwidth]{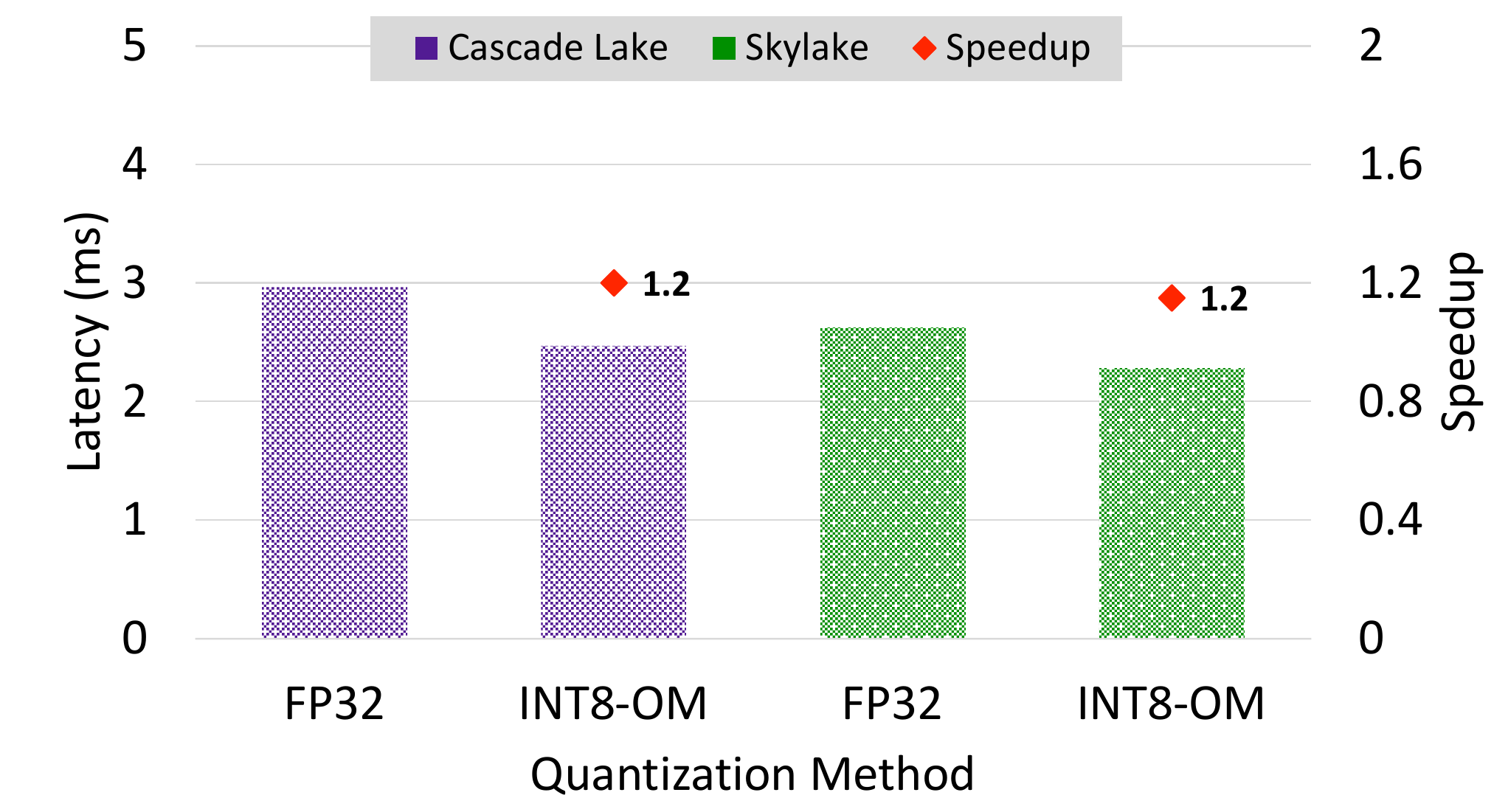}
         \caption{Single-stream}
     \end{subfigure}
     \hfill
     \begin{subfigure}[b]{.3\textwidth}
         \centering
         \includegraphics[width=\textwidth]{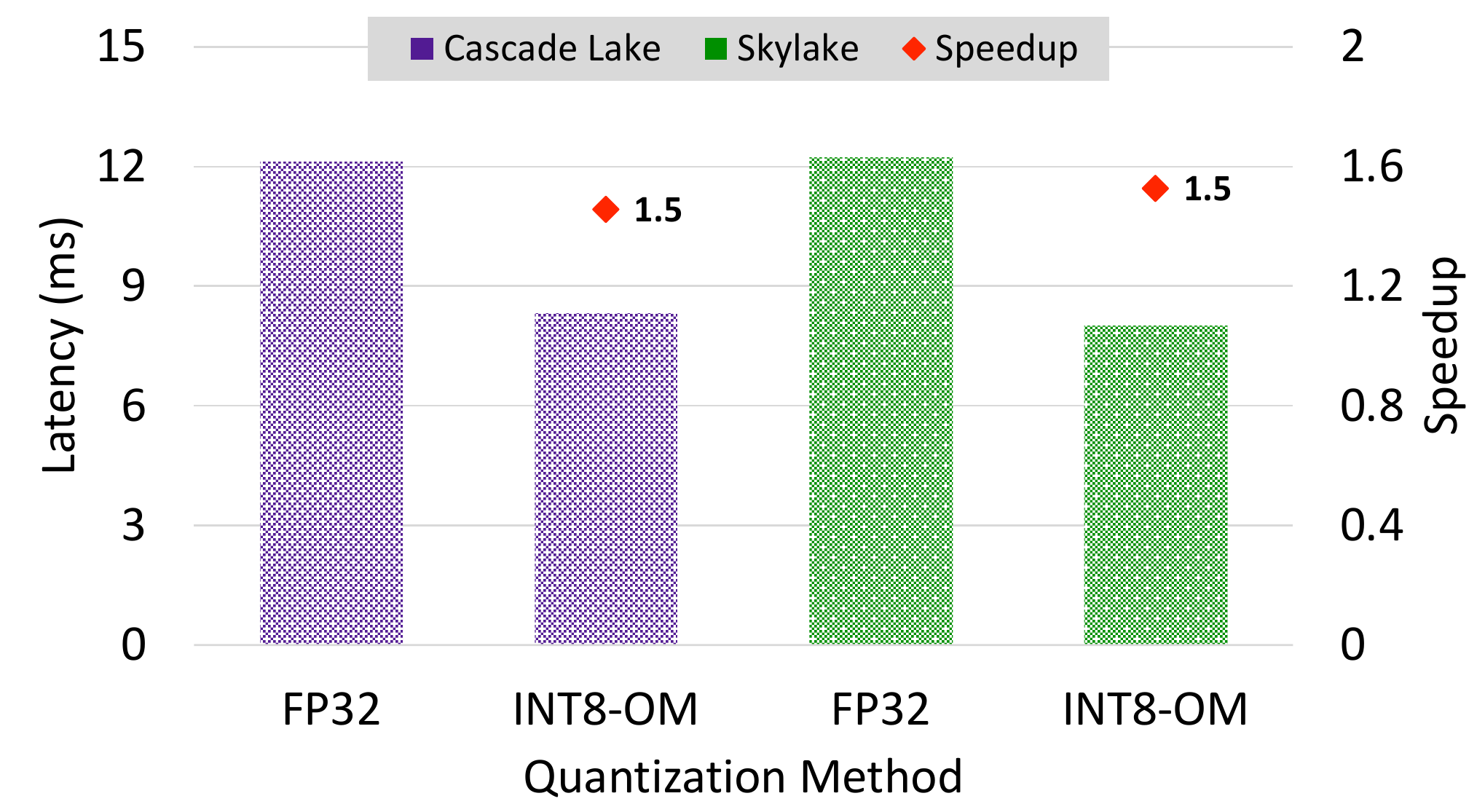}
         \caption{Multi-stream}
     \end{subfigure}
     \hfill
     \begin{subfigure}[b]{0.3\textwidth}
         \centering
         \includegraphics[width=\textwidth]{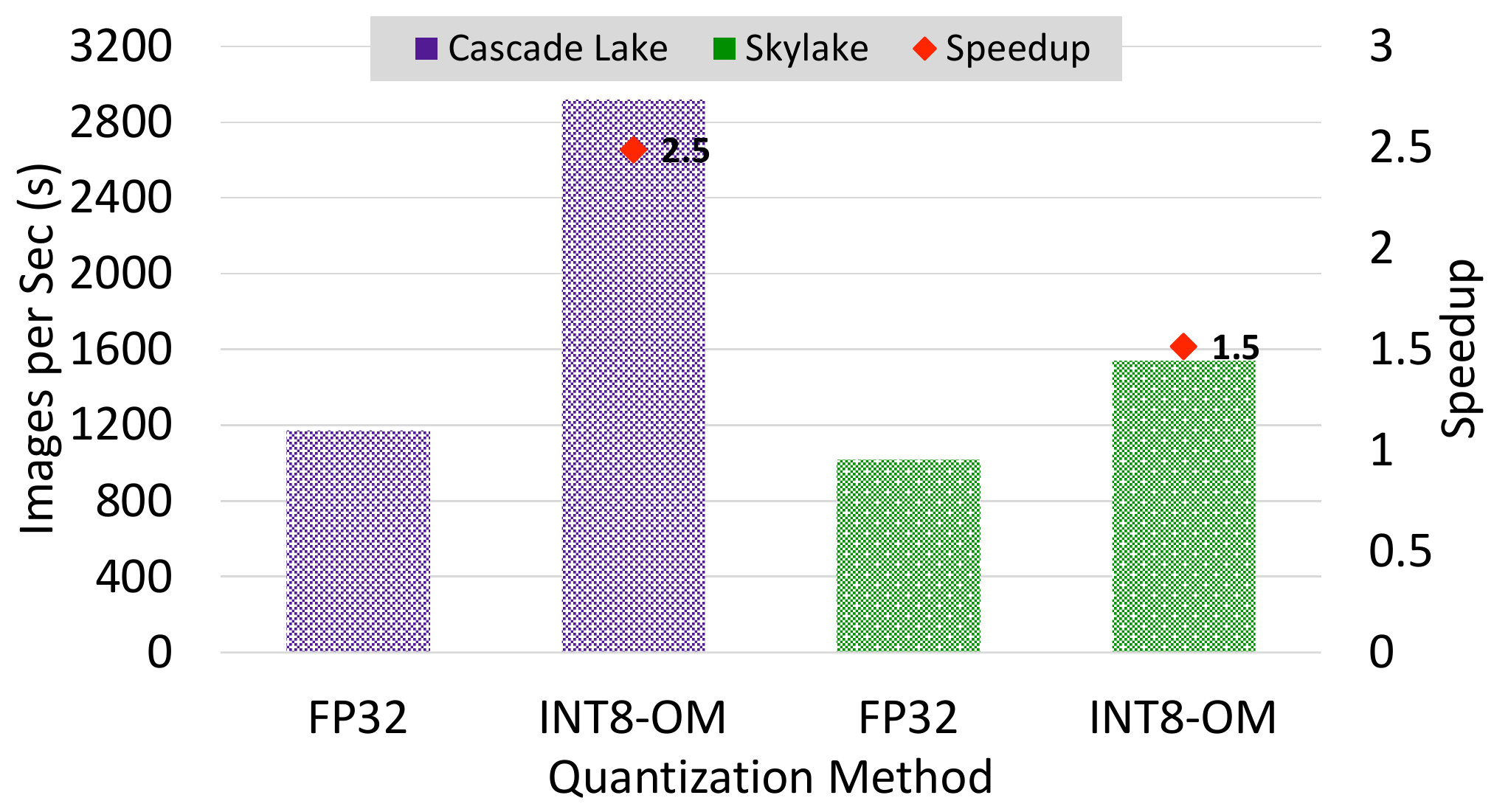}
         \caption{Offline}
         \label{fig:cascadelake-vs-skylake-offline}
     \end{subfigure}
        \caption{Inference performance of OpenVINO on Frontera (Cascade Lake processors) and RI2 (Skylake processors) using MLPerf Edge benchmarks. The model is MobileNetV2. Speedup is also plotted with diamonds on the {\tt y2} axis using the formula: $\frac{quantized\_performance}{FP32\_performance}$ for offline, $\frac{FP32\_performance}{quantized\_performance}$ for single/multi-stream.}
        \label{fig:cascadelake-vs-skylake}
\end{figure*}

Many vendors are now providing hardware support for accelerating inference tasks involving quantized weights/activations. In this sub-section, 
we demonstrate the benefits of using a newer generation of Intel CPU (Cascade Lake vs. Skylake) for the inference performance evaluation---single-stream latency, multi-stream latency, and offline scenarios---with the OpenVINO framework. This is depicted in Figure~\ref{fig:cascadelake-vs-skylake}, where Frontera and RI2 systems are equipped with Cascade Lake and Skylake processors, respectively. The main reason for better performance---especially for the 
offline scenario shown in Figure~\ref{fig:cascadelake-vs-skylake-offline} (see $2.5\times$ vs. $1.5\times$ speedup)---is that the Cascade Lake processors are equipped with AVX-512 Vector Neural Network Instructions (VNNI)\cite{VNNI} boosting INT8 operations.

%% file: text/related_work.tex
\section{Related Work}
\label{sec:related_work}

Quantization is a widely adopted method for edge device deep learning model inference. 
In \cite{Survey_of_quantization}, Gholami et al. conduct a survey of quantization methods. They state quantization could give benefits over multiple hardware devices like NVIDIA GPUs and ARM CPUs. Quantization results in higher power efficiency and performance, especially for edge devices. However, the study is a survey of existing quantization methods, hence, there are no numerical results in the survey.

In \cite{review_framworks}, Ulker et al. benchmark half-precision quantization on different devices with various state-of-the-art Deep Learning Frameworks. They provide detail on framework compatibility and indicate the best frameworks for each device model combination. They report the throughput and benefits of using half precision. However, no other quantization methods are further introduced, and the study has not covered frameworks targeting edge devices like TFLite. 

Efforts have been made in \cite{https://doi.org/10.48550/arxiv.2006.10226} to use compiler-based approaches to generate quantized models optimized for various platforms with different device types. However, the authors use quantized models as sanity checks for their compiler approach with limited quantization methods adopted.

In our work, we conduct a thorough analysis of multiple quantization methods in conjunction with popular deep-learning frameworks and hardware platforms from both the edge and high-end servers' worlds.

%% file: text/conclusion.tex
\section{Conclusions}
\label{sec:conclusion}
Quantization is a useful technique in DNN inference since it reduces memory footprint of AI models and improves performance without incurring accuracy loss. However, the diversity of edge devices and DNN frameworks makes it hard to adopt this technique and get the desired performance gains. In this paper, we evaluated several quantization methods of TFLite, Pytorch, ONNX, and OpenVINO on Intel Skylake, Intel Cascade Lake, and ARMv8 processors with MobileNetV2, DenseNet-121, and VGG-19. We utilized the methodology of the MLPerf Edge Inference benchmark with three scenarios---single-stream, multi-stream, and offline---to thoroughly understand the characteristic of quantization. The paper studied important quantization features including number format (like FP16 and INT8), symmetric vs. asymmetric, and static vs. dynamic approaches. We showed quantization can achieve up to $4.3\times$ times speedup compared to FP32. However, in the absence of instruction set support and/or algorithmic optimizations such as those adopted by FBGEMM, quantization can adversely impact the inference performance. 
In addition to the edge platform studied herein, we compared two generations of Intel processors (Skylake vs. Cascade Lake) to emphasize the effect of hardware and library support on quantization.
Overall, we highlighted the characteristics of quantization to help developers and researchers effectively adopt it in their particular configuration. 
In the future, we plan to study, evaluation, and characterize the impact of quantization on NVIDIA edge devices including AGX Orin using TensorRT inference framework.

%% file: text/acknowledgments.tex
\section{Acknowledgments}
\label{sec:ack}

This research is supported in part by NSF grants \#1818253, \#1854828, \#1931537, \#2007991, \#2018627, \#2112606, and XRAC grant \#NCR-130002.